\documentclass[sort&compress]{jfm}

\usepackage{graphicx}
\usepackage{newtxtext}
\usepackage{newtxmath}
\usepackage{natbib}
\usepackage{subcaption}

\usepackage{hyperref}
\usepackage[noabbrev,nameinlink]{cleveref}
\usepackage{algorithmic,algorithm}
\hypersetup{
    colorlinks = true,
    urlcolor   = blue,
    citecolor  = black,
}
\newcommand{\RomanNumeralCaps}[1]
\linenumbers

\title{Shapes optimising grand resistance tensor entries for a rigid body in a Stokes flow}

\author{Cl\'{e}ment Moreau\aff{1}\corresp{\email{cmoreau@kurims.kyoto-u.ac.jp}}, Kenta Ishimoto\aff{1},
  \and Yannick Privat\aff{2}\aff{3}}

\affiliation{\aff{1} Research Institute for Mathematical Sciences, Kyoto University, Kyoto 606-8501, Japan
\aff{2}IRMA, Universit\'e de Strasbourg, CNRS UMR 7501, Inria, 7 rue Ren\'e Descartes, 67084 Strasbourg, France
\aff{3}Institut Universitaire de France (IUF)}

\usepackage{amsmath,amssymb,mathtools}
\usepackage{graphicx}
\usepackage{overpic}
\usepackage{bm}
\usepackage{tikz}
\usetikzlibrary{hobby}
\usepackage{mathrsfs}
\usepackage{enumerate}

\renewcommand{\vec}[1]{\bm{#1}}

\newtheorem{proposition}{Proposition}

\begin{document}
\maketitle

\begin{abstract}
We investigate the optimal shapes of the hydrodynamic resistance of a rigid body set in motion in a Stokes flow. In this low Reynolds number regime, the hydrodynamic drag properties of an object are encoded in a finite number of parameters contained in the grand resistance tensor. Considering these parameters as objective functions to be optimised, we use calculus of variations techniques to derive a general shape derivative formula, allowing to specify how to deform the body shape to improve the objective value of any given resistance tensor entry. We then describe a practical algorithm for numerically computing the optimized shapes and apply it to several examples. Numerical results reveal interesting new geometries when optimizing the extra-diagonal inputs to the strength tensor, including the emergence of a chiral helical shape when maximising the coupling between the hydrodynamic force and the rotational motion. With a good level of adaptability to different applications, this work paves the way for a new analysis of the morphological functionality of microorganisms and for future advances in the design of microswimmer devices.

\end{abstract}

\begin{keywords}
low-Reynolds-number flows, hydrodynamic resistance, shape optimisation, Hadamard boundary variation method, augmented Lagrangian algorithm.
\end{keywords}

{\bf MSC Codes }  49Q10, 76D07.

\section{Introduction}
\label{sec:intro}

The interaction between solid objects and a surrounding fluid is at the heart of many fluid mechanics problems stemming from various fields such as physics, engineering and biology. Among other factors, the behaviour of such fluid-structure interaction systems is critically determined by the boundary conditions at the surface of the solid, but also by the geometry of the solid itself, or, more simply said, its shape. In this context, the research for some notion of shape optimality in the fluid-structure interaction is widespread, with the objective of understanding which shapes allow for optimal response from the fluid, typically involving energy-minimising criteria \citep{MR2567067}. 

At low Reynolds number, a regime occurring in particular at the microscopic scale where viscosity dominates on inertial effects, fluid dynamics are governed by the Stokes equations. These equations are linear and time-reversible -- an remarkable specificity compared to the more general Navier-Stokes equations, which makes fluid-structure interaction and locomotion at microscopic scale a peculiar world \citep{purcell1977life}. 

In particular, when considering the \textit{resistance problem} of a rigid body moving into a fluid at Stokes regime, a linear relationship holds between the motion of a body (translation and rotation) and the effects (forces and torques) it experiences. This relationship is materialised by the well-known \textit{grand resistance tensor}, which depends only, once a reference frame has been set, on the shape of the object and not on its motion on or the boundary conditions on fluid velocity. In other words, the hydrodynamic resistance properties of an object at low Reynolds number are intrinsic, contained in a finite number of parameters, which in turn are determined by its shape only. The question of which shapes possess maximal or minimal values for these resistance parameters then naturally arises, both from a theoretical fluid mechanics perspective, and as potential ways to explain the sometimes intriguing geometries of microorganisms \citep{yang2016staying,van2017determinants,lauga2020fluid, ryabov2021shape}. 

Optimal shapes for resistance problems have been tackled in previous studies. In particular, the minimal drag problem, which seeks the shape of fixed volume opposing the least hydrodynamic resistance to translation in a set direction, is well known and was solved in the 1970s, both analytically \citep{pironneau1973optimum} and numerically \citep{bourot1974numerical}. The characteristic rugby-ball shape resulting from this optimisation problem has then been used as a reference for many later works, among which we can cite the adaptations to two-dimesional and axisymmetric flows in \citet{richardson1995optimum,srivastava2011optimum}, linear elastic medium in \citet{zabarankin2013minimum}, or minimal drag for fixed surface in \citet{montenegro2015other}. These studies rely on symmetry properties for the minimal drag problem, and such methods fail to be immediately extended to solve the optimisation of the generic resistance problem, associated to other entries of the resistance tensor. 

Shape optimisation in microhydrodynamics has also been widely carried in the context of microswimmer locomotion. Notable works include \citet{quispe2019geometry}, where the best pitch and cross-section for efficient magnetic swimmers is numerically and experimentally discussed, and \citet{fujita2001optimum,ishimoto2016hydrodynamic, berti2021shapes}, where parametric optimisation is conducted to find the best geometry for flagellated microswimmers. Efficient shapes for periodic swimming strokes and ciliary locomotion are addressed in \citet{vilfan2012optimal,daddi2021optimal}. 

However, in all of these studies, restrictive assumptions are made on the possible shapes, with the optimisation being carried on a few geometrical parameters and not on a general space of surfaces in 3D. Another approach, allowing to explore a wider class of shapes than with parametric optimisation, is based on the use of shape derivatives: a generalisation of the notion of derivative, which yields a perturbation function of a domain in a descent direction. This method however requires caution regarding the regularity assumptions on the boundaries of the domains involved. Other popular methods for shape optimisation in structural mechanics include density methods, in which the characteristic function of a domain is replaced by a density function -- we mention in particular the celebrated SIMP method \citep{bendsig,borvall,evgrafov}, and the level set method, \citep{osher1988fronts,sethian2000structural,ajt,wang} which can handle changes of topology. Obtaining efficient numerical algorithms to apply these analytical methods to find optimal shapes is also challenging: one must be able to handle both a decrease of the objective function, while avoiding that the numerical representation becomes invalid (for example because of problems related to the mesh or to changes in the topology of the shapes considered). 

In the context of low-Reynolds number fluid mechanics, variational techniques and shape derivatives are notably used by \citet{Walker2013} and \citet{keaveny2013optimization} to carry the optimisation of the torque-speed mobility coefficient in the context of magnetically propelled swimmers, for a shape constrained to be a slender curved body, yielding helicoidal folding. However, this study is also focused on a restricted class of shapes, characterised by a one-dimensional curve, and a single entry of the resistance tensor or energy dissipation criteria. To the best of the authors' knowledge, the systematic theoretical or numerical study of the coefficients of the grand resistance tensor has not yet been carried out. 

Hence, as the principal aim of this paper, we will provide a general framework of shape optimisation for this type of problem, and show that the optimisation of any entry of the resistance tensor amounts to a single, simple formula for the shape derivative, which depends on the solution of two Stokes problems whose boundary conditions depend on the considered entry. We then describe an algorithm to numerically implement the shape optimisation and display some illustrative examples. 




\section{Problem statement}
\label{sec:problem}

\subsection{Resistance problem for a rigid body in Stokes flow}

\begin{figure}
\begin{center}
\begin{overpic}[height=5cm]{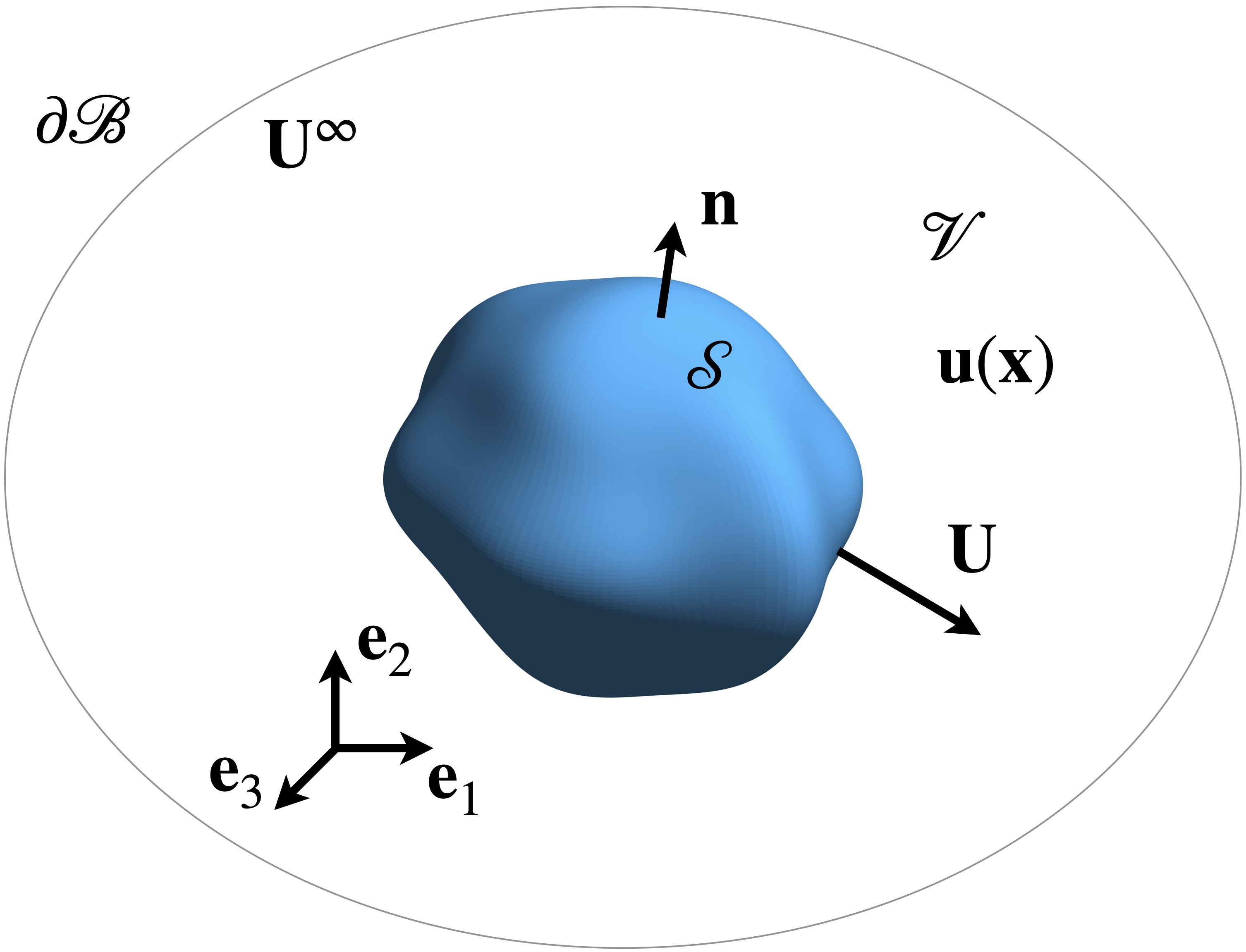}
\put(26,12){$O$}
\end{overpic}
$\quad$
\includegraphics[height=5cm]{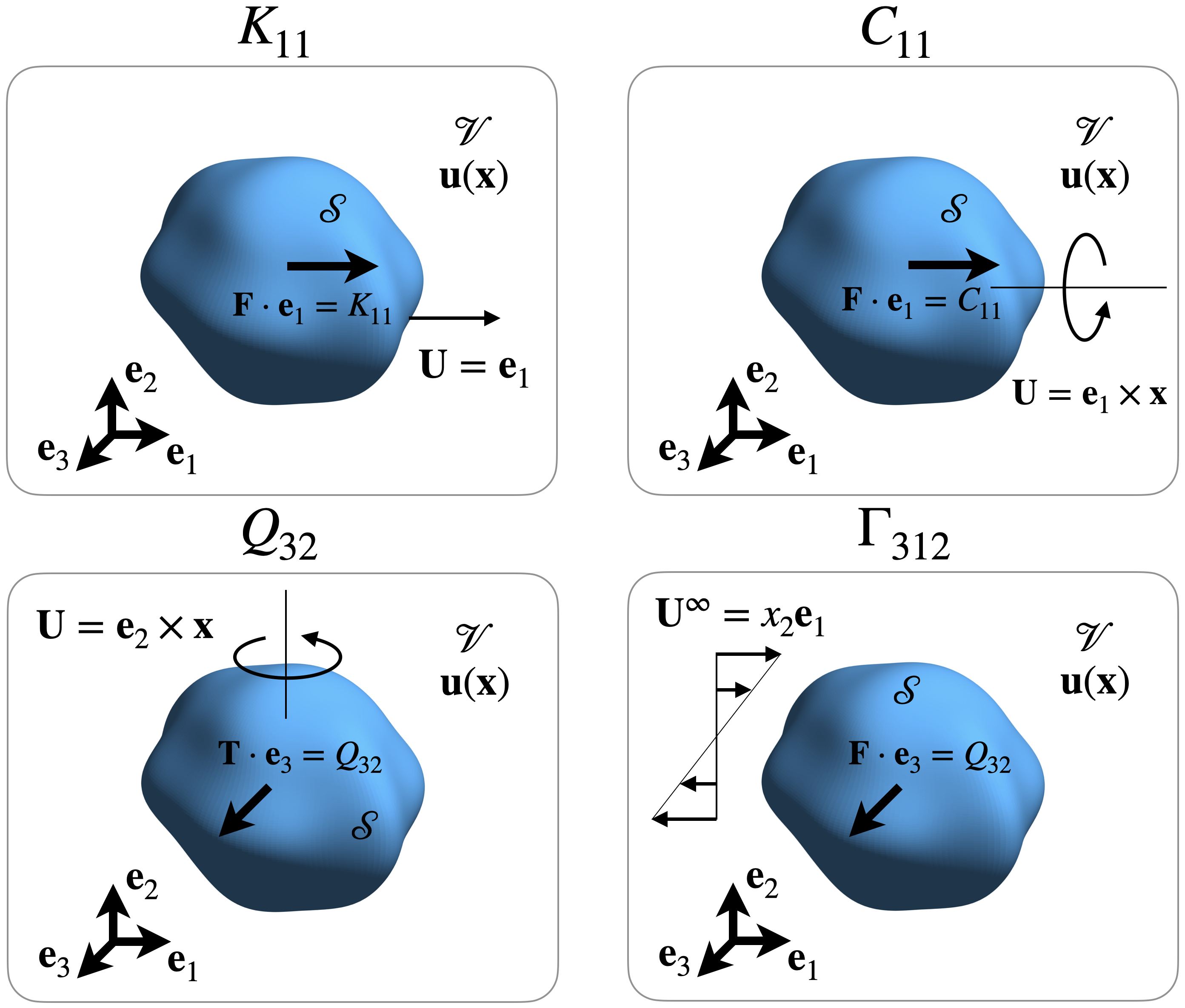}
\end{center}
\caption{Problem setup: a rigid body in a Stokes flow. A diagram of the general Stokes problem \eqref{eq:stokes} can be seen on the left of the figure. The panels on the right-hand side show examples of resistance problems associated to selected entries of the grand resistance tensor. For instance, for $K_{11}$ (top left), one sets the motion of the object to a unitary translation in the direction $\vec{e}_1$, and then $K_{11}$ may be obtained as the component along $\vec{e}_1$ of the total drag force $\vec{F}$ exerted on the object. The other coefficients shown on the other panels are analogously obtained by using the appropriate boundary conditions and drag force or torque shown on the figure. \label{fig:scheme}}
\end{figure}

We consider a rigid object set in motion into an incompressible fluid with viscosity $\mu$ at low Reynolds number, with coordinates $\vec{x}$ expressed in the fixed lab frame $(O,\vec{e}_1,\vec{e}_2,\vec{e}_3)$, as shown on the left panel of \Cref{fig:diagram}. The object's surface is denoted by $\mathscr{S}$ and we assume that the fluid is contained in a bounded domain $\mathscr{B}$, thus occupying a volume $\mathscr{V}$ having $\partial \mathscr{V}= \mathscr{S}\cup \partial \mathscr{B}$ as boundary. Such a boundedness assumption ensures that the solutions of the fluid equations are well-defined and that the computations that will be performed on them throughout the paper are rigorously justified (see Appendix \ref{append:diff} and textbooks referred therein), although we assume the outer boundary $\partial \mathscr{B}$ to be sufficiently far from the object for its effect on hydrodynamic resistance to be negligible. At the container boundary $\partial \mathscr{B}$, we consider a uniform, linear background flow $\vec{U}^\infty$, broken down into translational velocity vector $\vec{Z}$, rotational velocity vector $\vec{\Omega}^{\infty}$ and rate-of-strain (second-rank) tensor $\vec{E}^{\infty}$ components as follows:
\begin{equation}\label{def:Uinfty}
\vec{U}^\infty=\vec{Z} + \vec{\Omega}^{\infty} \times \vec{x} + \vec{E}^{\infty} \vec{x}.
\end{equation}
Similarly, the object's rigid motion velocity field is simply described by 
\begin{equation}\label{def:U}
\vec{U} = \vec{Z} + \vec{\Omega} \times \vec{x},
\end{equation}
with $\vec{U}$ and $\vec{\Omega}$ denoting its translational and rotational velocities. Having set as such the velocities at the boundary of $\mathscr{V}$ defines a boundary value problem for the fluid velocity field $\vec{u}$ and pressure field $p$, which satisfy the Stokes equations:
\begin{equation}
\left \{
\begin{array}{ll}
\mu \Delta \vec{u} - \nabla p = \vec{0} & \text{in }\mathscr{V}, \\
\nabla \vec{\cdot} \vec{u} = 0& \text{in }\mathscr{V}, \\
\vec{u} = \vec{U} & \text{on } \mathscr{S}, \\
\vec{u} = \vec{U}^{\infty}  & \text{on } \partial \mathscr{B}.
\end{array}
\right .
\label{eq:stokes}
\end{equation}
From the solution of this Stokes problem with set boundary velocity, one can then calculate the hydrodynamic drag (force $\vec{F}^h$ and torque $\vec{T}^h$) exerted by the moving particle to the fluid, \textit{via} the following surface integrals formulae over $\mathscr{S}$: 
\begin{align}
    \vec{F}^h & = - \int_{\mathscr{S}} \vec{\sigma}[\vec{u},p] \vec{n} \mathrm{d} \mathscr{S}, \\
    \vec{T}^h & = - \int_{\mathscr{S}} \vec{x} \times \left ( \vec{\sigma}[\vec{u},p] \vec{n} \right ) \mathrm{d} \mathscr{S}.
    \label{eq:FT}
\end{align}
In \Cref{eq:FT}, $\vec{n}$ is the normal to $\mathrm{d} \mathscr{S}$ pointing outward to the body (see Fig. \ref{fig:scheme}), and $\vec{\sigma}$ is the stress tensor, defined as
\begin{equation}
    \vec{\sigma} [\vec{u},p] = - p \vec{I} + 2 \mu \vec{e}[\vec{u}],
\end{equation}
in which $\vec{I}$ denotes the identity tensor and $\vec{e}[\vec{u}]$ is the rate-of-strain tensor, given by
\begin{equation}
    \vec{e}[\vec{u}] =  \frac{1}{2} \left ( \nabla \vec{u} + (\nabla \vec{u} )^T \right ).
\end{equation}
Finding this way the hydrodynamic drag for a given velocity field is called the \textit{resistance problem} -- as opposed to the \textit{mobility problem} in which one seeks to find the velocity generated by a given force and torque profile on the boundary. 


\subsection{Grand resistance tensor}

In addition to Equations, \eqref{eq:FT}, a linear relationship between $(\vec{F}^h,\vec{T}^h)$ and $(\vec{U},\vec{U}^{\infty})$ can be derived from the linearity of the Stokes equation \citep[see][Chapter 5]{Kim2005}:
\begin{equation} 
\begin{pmatrix} 
\vec{F}^h \\ \vec{T}^h \\ \vec{S} \end{pmatrix} = \mathsfbi{R} \begin{pmatrix} \vec{Z} - \vec{Z}^{\infty}  \\ \vec{\Omega}- \vec{\Omega}^{\infty} \\ - \vec{E}^{\infty} \end{pmatrix} = \begin{pmatrix} \mathsfbi{K} & \mathsfbi{C} & \vec{\Gamma} \\ \tilde{\mathsfbi{C}} & \mathsfbi{Q} & \vec{\Lambda}  \\  \tilde{\vec{\Gamma}} & \tilde{\vec{\Lambda}} & \mathsfbi{M} \end{pmatrix} \begin{pmatrix} \vec{Z} - \vec{Z}^{\infty}  \\ \vec{\Omega} - \vec{\Omega}^{\infty} \\ - \vec{E}^{\infty} \end{pmatrix}.
\label{eq:GRT}
\end{equation}
The stresslet $\vec{S}$, defined as
\begin{equation}
     \vec{S} = \frac{1}{2} \int_{\mathscr{S}} (\vec{x} \vec{\cdot}  \vec{\sigma}[\vec{u},p] \vec{n}^T +  \vec{\sigma}[\vec{u},p] \vec{n}  \vec{\cdot} \vec{x}^T ) \mathrm{d} \mathscr{S}, 
\end{equation}
appears on the right-hand side of Equation \eqref{eq:GRT} and is displayed here for the sake of completeness, though we will not be dealing with it in the following. 

The tensor $\mathsfbi{R}$, called the grand resistance tensor, is symmetric and positive definite. As seen in Equation \eqref{eq:GRT}, it may be written as the concatenation of nine tensors, each accounting for one part of the force-velocity coupling. The second-rank tensors $\mathsfbi{K}$ and $\mathsfbi{C}$ represents the coupling between hydrodynamic drag force and, respectively, translational and rotational velocity. Similarly, $\tilde{\mathsfbi{C}}$ and $\mathsfbi{Q}$ are second-rank tensors coupling hydrodynamic torque with translational and rotational velocity. Note that, by symmetry of $\mathsfbi{R}$, $\mathsfbi{K}$ and $\mathsfbi{Q}$ are symmetric and one has $\mathsfbi{C}^T = \tilde{\mathsfbi{C}}$. Further, $\vec{\Gamma}$, $\tilde{\vec{\Gamma}}$, $\vec{\Lambda}$, and $\tilde{\vec{\Lambda}}$ are third-rank tensors accounting for coupling involving either the shear part of the background flow or the stresslet, and $\mathsfbi{M}$ is a fourth-rank tensor representing the coupling between the shear and the stresslet, with similar properties deduced from the symmetry of $\mathsfbi{R}$.  

An important property of the grand resistance tensor is that it is independent of the boundary conditions associated to a given resistance problem. In other words, for a given viscosity $\mu$ and once fixed a system of coordinates, the grand resistance tensor $\mathsfbi{R}$ depends only on the shape of the object, \textit{i.e.} its surface $\mathscr{S}$. A change of coordinates or an affine transformation applied to $\mathscr{S}$ modifies the entries of $\mathsfbi{R}$ through standard linear transformations. For that reason, here we fix a reference frame once and for all and carry the shape optimisation within this frame; which means in particular that we distinguish shapes that do not overlap in the reference frame, even if they are in fact identical after an affine transformation. 

With these coordinates considerations aside, we can argue that the grand resistance tensor constitutes an intrinsic characteristic of an object; and all the relevant information about the hydrodynamic resistance of a certain shape is carried in the finite number of entries in $\mathsfbi{R}$. While these entries can be obtained by direct calculation in the case of simple geometries, in most cases their value must be determined by solving a particular resistance problem and using Equations $\eqref{eq:FT}$. For example, to determine $K_{ij}$, one can set $\vec{U}$ as unit translation along $\vec{e}_j$, $\vec{U} = \vec{e}_j$. Then Equation \eqref{eq:GRT}, combined with $\eqref{eq:FT}$, gives
\begin{equation}
K_{ij} = \vec{F}^h \vec{\cdot} \vec{e}_i = - \int_{\mathscr{S}} (\vec{\sigma}[\vec{u},p] \vec{n})\vec{\cdot} \vec{e}_i \mathrm{d} \mathscr{S}.
\end{equation}
The same strategy can be applied for other entries of $\mathsfbi{R}$, setting appropriate boundary conditions $\vec{U}$ and $\vec{U}^{\infty}$ in the Stokes equation and calculating the appropriate projection of $\vec{F}^h$ or $\vec{T}^h$ along one of the basis vectors. Figure \ref{fig:scheme} displays a few illustrative examples. In fact, let us define the generic quantity $J_{\vec{V}}$ as the surface integral
\begin{equation}
    J_{\vec{V}}(\mathscr{S}) = - \int_{\mathscr{S}} (\vec{\sigma}[\vec{u},p] \vec{n})\vec{\cdot} \vec{V} \mathrm{d} \mathscr{S}.
    \label{eq:J}
\end{equation}
Then, judicious choices of $\vec{U}$, $\vec{U}^{\infty}$ and $\vec{V}$, summarised in Table \ref{table:1}, allow to obtain any coefficient of the grand resistance tensor from formula \eqref{eq:J}.

\begin{table}
  \begin{center}
\def~{\hphantom{0}}
  \begin{tabular}{c c c c}
      $\qquad J_{\vec{V}} \qquad$  & $\qquad \vec{U} \qquad$   &  $\qquad \vec{V} \qquad$ & $\qquad \vec{U}_{\infty} \qquad$ \\[3pt] 
       $K_{ij}$   & $\vec{e}_j$ & $\vec{e}_i$ & $\vec{0}$ \\[2pt]
       $C_{ij}$   & $\vec{e}_j \times \vec{x}$ & $\vec{e}_i$ & $\vec{0}$ \\[2pt]
       $\tilde{C}_{ij}$  & $\vec{e}_j$ & $\vec{e}_i \times \vec{x}$ & $\vec{0}$\\[2pt]
       $Q_{ij}$   & $\vec{e}_j \times \vec{x}$ & $\vec{e}_i \times \vec{x}$ & $\vec{0}$\\[2pt]
       $\Gamma_{ijk}$ & $\vec{0}$ & $\vec{e}_i$ & $x_k \vec{e}_j$ \\[2pt]
       $\Lambda_{ijk}$ & $\vec{0}$ & $\vec{e}_i \times \vec{x}$ & $x_k \vec{e}_j$ \\
  \end{tabular}
  \caption{Entries of the grand resistance tensor associated to $J$ with respect to the choice of $\vec{U}$, $\vec{V}$ and $\vec{U}^{\infty}$.}
  \label{table:1}
  \end{center}
\end{table}

Of note, for the determination of the coefficients lying on the diagonal of $\mathsfbi{R}$, another relation involving power instead of hydrodynamic force is sometimes found \citep{Kim2005}. Indeed, the energy dissipation rate $\epsilon$ is defined as $\epsilon = \int_{\mathscr{V}} 2 \mu \vec{e}[u] : \vec{e}[u] \mathrm{d} \mathscr{V}$. In the case of the translation $\vec{U}$ of a rigid body, one also has $\epsilon = \vec{F} \vec{\cdot} \vec{U}$. Then, to determine for instance $K_{11}$, one sets $\vec{U} = \vec{e}_1$ as described above and obtains
\begin{equation}
    K_{11} = \int_{\mathscr{V}} 2 \mu e_{1j} e_{1j} \mathrm{d}\mathscr{V}.
\end{equation}
This last expression yields in particular the important property that the diagonal entries of $\mathsfbi{R}$ are positive. Nonetheless, in the following we will prefer the use of formula \eqref{eq:J} that conveniently works for both diagonal and extradiagonal entries. 

\subsection{Towards a shape optimisation framework}

Seeing $J_{\vec{V}}$ as a functional depending on the surface $\mathscr{S}$ of the object, we will now seek to optimise the shape $\mathscr{S}$ with $J_{\vec{V}}$ as an objective function; in other terms, we want to optimise one of the parameters accounting for the hydrodynamic resistance of the object. 
As is usually done in shape optimisation, it is relevant in our framework to add some constraint on the optimisation problem. This is both motivated by our wish to obtain relevant and non-trivial shapes (e.g. a shape occupying the whole computational domain), but also to model manufacturing constraints. In this domain, there are multiple choices. In the following, we will focus on the standard choice: $|\mathscr{V}|=V_0$ for some positive parameter $V_0$, where $\mathscr{V}$ stands for the domain enclosed by $\mathscr{S}$ whereas $|\mathscr{V}|$ denotes its volume.

The {\it generic} resulting shape optimisation problem we will tackle in what follows hence reads:
\begin{equation}\label{SOPgen-min}
\min_{\mathscr{S}\in \mathscr{O}_{ad,V_0}} J_{\vec{V}}(\mathscr{S}),
\end{equation}
where $\mathscr{O}_{ad,V_0}$ denotes the set of all connected domains $\mathscr{V}$ enclosed by $\mathscr{S}$ such that $|\mathscr{V}|=V_0$. Of note, when performing optimisation in practice, we will also occasionally consider $\max J_{\vec{V}}(\mathscr{S})$ instead of $\min J_{\vec{V}}(\mathscr{S})$ in \eqref{SOPgen-min}, which is immaterial to the following analysis as it amounts to replacing $J_{\vec{V}}(\mathscr{S})$ by $-J_{\vec{V}}(\mathscr{S})$.

Throughout the rest of this paper, we will not address the question of the existence of optimal shapes, i.e. the existence of solutions for the above problem. In the context of shape optimisation for fluid mechanics, few qualitative analysis questions have been solved so far. Let us mention for instance \citet{MR2601075} about some progress in this field, whether it is about questions of existence or qualitative analysis of optimal shapes. Whilst it may appear as purely technical, these fundamental questions can have very tangible impact on the actual optimal shapes; for instance, necessary regularity assumptions influence the decisions to be made for numerical implementation, and a reckless choice of admissible shapes space in which the optimisation problem does not have any solution may thus lead to ``miss'' the minimiser.

Nevertheless, we put these questions aside in this paper, in order to focus on the presentation of an efficient optimisation algorithm based on the use of shape derivatives, as well as the numerical results obtained and their interpretation.

We will hence use the framework of shape derivative calculus, which allows to consider very general shape deformations, independent of any shape parametrisation, and a global point of view on the objective function; this notably constitutes a progress with respect to previous studies focusing on one particular entry of the resistance tensor. The following section is devoted to laying out the mathematical tools required to address the hydrodynamic shape optimisation problem. 




\section{Analysis of the shape optimisation problem}

\subsection{Theoretical framework}\label{sec:theoFramSO}

In this section, we introduce the key concepts of domain variation and shape gradient that we need to introduce the main results of this paper, as well as a practical optimisation algorithm. 

An important point that must be considered throughout this study is shape regularity. In order for the presented calculations and the equations involved to be valid and understood in their usual sense, sufficient regularity of the surfaces involved must be imposed; however, it must be kept in mind that the need to consider regular shapes might prevent some optimal shapes from being found if they possess ridges or sharp corners. In addition, the discretisation step required when moving to the numerical implementation also raises further discussions about the regularity of shapes. 

For the sake of readability, the mathematical framework, especially with respect to shape regularity and the other functional spaces involved, is kept to a minimal level of technicality in the body of the paper and further details and discussions are provided in the appendix \ref{append:diff}.

\begin{figure}
\begin{center}
\includegraphics[width=8cm]{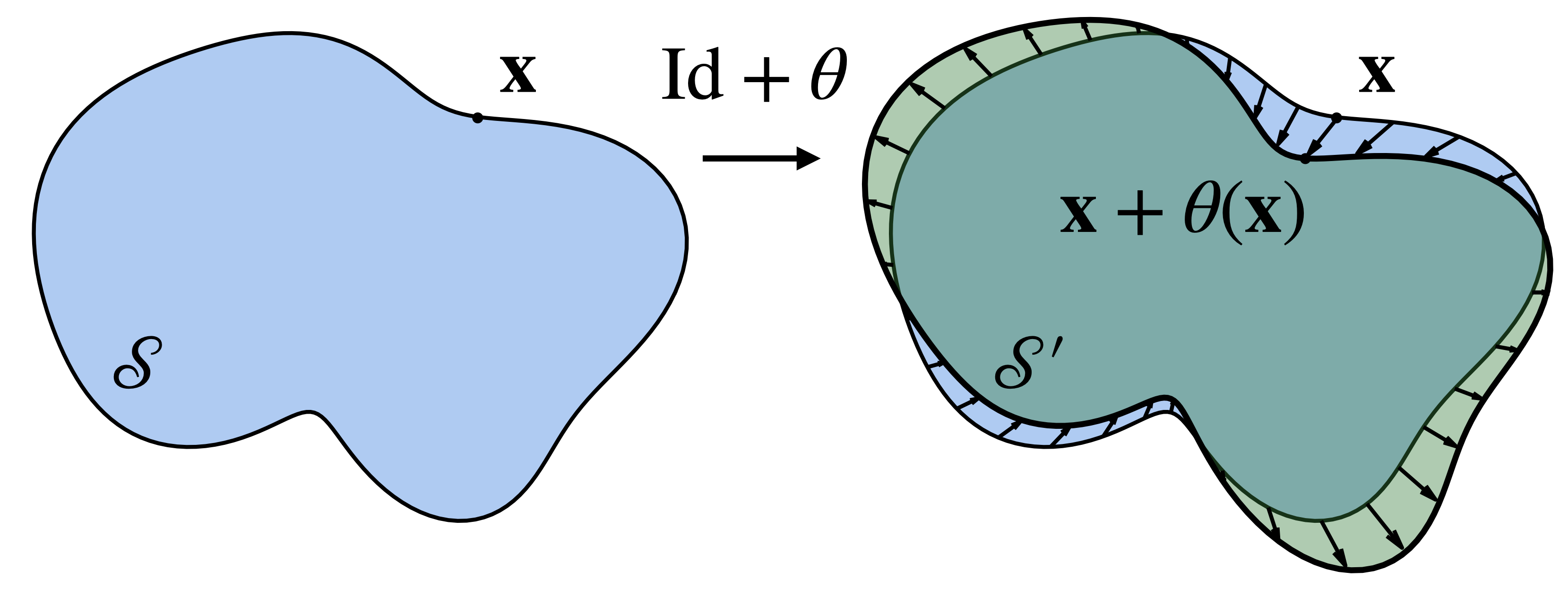}
\caption{Shape optimisation principle: the surface $\mathscr{S}$ of the body is deformed with respect to a certain vector field $\vec{\theta}$, such that the deformed shape $\mathscr{S}_{\vec{\theta}} = (\mathrm{Id} + \vec{\theta}) (\mathscr{S})$ improves the objective, \textit{i.e.} satisfies $J(\mathscr{S}_{\vec{\theta}})<J(\mathscr{S})$.}
\label{fig:diagram}
\end{center}
\end{figure}

Formally, deforming a shape can be done by defining a deformation vector field $\vec{\theta}~:~\mathbb{R}^3~\rightarrow~\mathbb{R}^3$ within the domain $\mathscr{B}$. This vector field will be assumed to belong to a set $\vec{\Theta}$ of so-called ``admissible'' fields, that are smooth enough to preserve the regularity of the shape. A discussion on the exact choice of set $\vec{\Theta}$ can be found in the Appendix. 

Applying this deformation vector field $\vec{\theta}$ to the surface of the shape $\mathscr{S}$ yields a new shape $\mathscr{S}_{\vec{\theta}}= (\mathrm{Id} + \vec{\theta}) (\mathscr{S})$  (see figure \ref{fig:diagram}), where $\mathrm{Id}$ denotes the identity operator corresponding to no deformation. This operation is called a \textit{domain variation}. The fundamental question at the heart of all shape optimisation algorithms is the search for a "good" vector field $\vec{\theta}$ chosen so that $\mathscr{S}_{\vec{\theta}}$ satisfies the constraints of the problem but also so that the objective function decreases, ideally strictly -- but most methods only guarantee the inequality
$J(\mathscr{S}_{\vec{\theta}})\leq J(\mathscr{S})$. In the terminology of optimisation, such a deformation vector field will be called a \textit{descent direction}, according to the inequality above. In numerical optimisation, descent methods are expected to bring the shape towards a local optimum for the objective criterion. 

To this end, following the so-called Hadamard boundary variation framework which considers a small change of functional by perturbing shape geometry as featured in \citet{allaire2007conception} and \citet{HENROTPIERRE}, to which we refer the interested reader for a detailed theory of shape optimisation, we introduce the fundamental notion of \textit{shape derivative}, characterising the variation of the criterion for an infinitesimal deformation of the domain. More precisely, for a given shape $\mathscr{S}$ and $\vec{\theta} \in \vec{\Theta}$, the \textit{shape derivative of $\mathscr{S}$ in the direction $\vec{\theta}$} is denoted by $\langle \mathrm{d} J(\mathscr{S}),\bm{\theta}\rangle$ and defined as the first order term in the expansion
\begin{equation}
J(\mathscr{S}_{\vec{\theta}}) = J(\mathscr{S}) + \langle \mathrm{d} J(\mathscr{S}),\bm{\theta}\rangle + o(\bm{\theta})
\quad
\text{where } o(\bm{\theta}) \to 0 \text{ as } \bm{\theta} \to 0.
\label{eq:shape-diff}
\end{equation}
For additional explanations on the precise meaning of such an expansion, we refer to Appendix~\ref{append:funSpace}.

In particular, the \textit{shape derivative of $J$ at $\mathscr{S}$ in the direction $\vec{\theta}$} can be computed through the directional derivative
\begin{equation}
    \langle \mathrm{d} J(\mathscr{S}),\bm{\theta}\rangle = \lim_{\varepsilon \rightarrow 0} \frac{J((\mathrm{Id} + \varepsilon \vec{\theta}) (\mathscr{S}))- J(\mathscr{S})}{\varepsilon},
\end{equation}
from which we recover \Cref{eq:shape-diff}. Of note, the bracket notation for the shape derivative refers to the fact that the application $\vec{\theta} \mapsto \langle \mathrm{d} J(\mathscr{S}),\bm{\theta}\rangle$ is a linear form from $\vec{\Theta}$ to $\mathbb{R}$, which itself stems out from the differential $\mathrm{d}J (\mathscr{S})$ at $\vec{\theta} = 0$ of the domain variation application
\begin{equation}
\begin{array}{r c l}
    \vec{\Theta} & \rightarrow & \mathbb{R} \\
    \vec{\theta} &  \mapsto & J(\mathscr{S}_{\vec{\theta}}).
\end{array}
\end{equation}

The expression of the shape derivative in Equation \eqref{eq:shape-diff} suggests that the deformation $\vec{\theta}$ should be chosen such that $\langle \mathrm{d} J(\mathscr{S}),\bm{\theta}\rangle$ is negative, effectively decreasing the objective criterion at first order. A classical strategy to achieve this goal \citep[see][Chapter~6]{allaire2007conception} consists in deriving an explicit and workable expression of the shape derivative as a surface integral of the form 
\begin{equation}
    \langle \mathrm{d} J(\mathscr{S}),\bm{\theta}\rangle = \int_{\mathscr{S}} G(\vec{x}) \vec{\theta} \vec{\cdot} \vec{n} \mathrm{d} \mathscr{S}(\vec{x}),
    \label{eq:shape-grad}
\end{equation}
where $G$ is a function called \textit{shape gradient} of the involved functional. Such a rewriting is in general possible for generic cost functions (according to the structure theorem, see e.g. \cite[Section~5.9]{HENROTPIERRE}), but usually requires some work, and involves the determination of the adjoint of a linear operator.
Once an expression of type \eqref{eq:shape-grad} has been obtained, it is then easy to prescribe the descent direction such that the shape derivative is negative, by choosing for instance $\theta(\vec{x}) = -G(\vec{x}) \vec{n}$, or less straightforward expressions yielding suitable properties; see section \ref{sec:descent} for further discussion. 

\subsection{The shape derivative formula for problem \eqref{SOPgen-min}}

The calculation of the shape gradient through the derivation of a formula like Equation \eqref{eq:shape-grad} for the minimisation problem \eqref{SOPgen-min} is the main result of this theoretical section, which is displayed in Proposition \ref{prop:diff} below. In order to state this result, let us introduce the pair $(\vec{v},q)$ playing the role of \textit{adjoint states} for the optimisation problems we will deal with, as the unique solution of the Stokes problem
\begin{equation}
\left \{
\begin{array}{ll}
\mu \Delta \vec{v} - \nabla q = \vec{0} & \text{in }\mathscr{V} \\
\nabla \vec{\cdot} \vec{v} = 0& \text{in }\mathscr{V} \\
\vec{v} =\vec{V} & \text{on } \mathscr{S}, \\
\vec{v} = \vec{0} & \text{on } \partial \mathscr{B},
\end{array}
\right .
\label{eq:stokesAdj}
\end{equation}

Then, one can express the shape derivative and shape gradient with respect to the solution of the resistance problem \eqref{eq:stokes} and the adjoint problem \eqref{eq:stokesAdj}:

\noindent \begin{minipage}{\textwidth}
\begin{proposition}\label{prop:diff} The functional $J_{\vec{V}}$ is shape differentiable. Furthermore, for all $\bm{\theta}$ in $\Theta$, one has
\begin{equation}
 \langle dJ_{\vec{V}}(\mathscr{S}),\bm{\theta}\rangle = 2\mu \int_{\mathscr{S}}  \big ( \vec{e}[\vec{u}]:\vec{e}[\vec{v}] -\vec{e}[\vec{U} ] : \vec{e} [ \vec{v} ] - \vec{e} [\vec{u} ] : \vec{e} [ \vec{V} ] \big ) (\vec{\theta}\vec{\cdot}\vec{n}) \mathrm{d} \mathscr{S}, 
 \label{eq:shape_grad}
\end{equation}
and the shape gradient $G$ is therefore given by
$$
G = 2\mu  \big ( \vec{e}[\vec{u}]:\vec{e}[\vec{v}] -\vec{e}[\vec{U} ] : \vec{e} [ \vec{v} ] - \vec{e} [\vec{u} ] : \vec{e} [ \vec{V} ] \big ).
$$
\end{proposition}
\end{minipage}

Of particular note, if we assume moreover that $ \vec{U}$ and $ \vec{V}$ are such that
$$
\vec{e}[\vec{U}]= \vec{e}[ \vec{V}]= \vec{0} \text{ in }\mathscr{V},
$$
which is trivially true for all the relevant choices of $\vec{U}$ and $\vec{V}$ displayed in Table \ref{table:1} -- and more generally for any linear flow and rigid body motion -- then the shape gradient simply becomes
\begin{equation}
 G = 2\mu  \vec{e}[\vec{u}]:\vec{e}[\vec{v}],
\end{equation}
which is the expression we will use later on when implementing the shape optimisation algorithm.

\subsection{Proof of Proposition \ref{prop:diff}}

To compute the shape gradient of the functional $J_{\vec{V}}$, which is expressed as a surface integral, a standard technique \citep[see][Chapter~5]{HENROTPIERRE} first consists in rewriting it under volumetric form. 

Let us multiply the main equation of \eqref{eq:stokes} by $\vec{v}$ and then integrate by parts\footnote{
In what follows, we will often use the following integration by parts formula, well adapted to the framework of fluid mechanics: let $\vec{u}$ and $\vec{v}$ denote two vector fields; then,
$
2\int_{\mathscr{V}} \vec{e} [\vec{v}]:\vec{e} [\vec{u}] \mathrm{d} {\mathscr{V}}=-\int_{\mathscr{V}} (\Delta \vec{v}+\nabla (\nabla \vec{\cdot} \vec{v}))\vec{\cdot} \vec{u}\mathrm{d}\mathscr{V}+2\int_{\partial \mathscr{V}}\vec{e} [\vec{v}]\vec{n}\vec{\cdot} \vec{u}\mathrm{d} \mathscr{S}.
$
}. One thus gets
$$
2\mu \int_{\mathscr{V}} \vec{e}[\vec{u}]:\vec{e}[\vec{v}]\mathrm{d} {\mathscr{V}}-\int_{\mathscr{V}} p\nabla \vec{\cdot} \vec{v}\mathrm{d} {\mathscr{V}}-2\int_{\partial \mathscr{V}}\vec{\sigma}[\vec{u},p]\vec{n}\vec{\cdot} \vec{v}\mathrm{d} \mathscr{S}=0
$$
By plugging the boundary conditions into this equality, one gets
\begin{equation}
-J_{\vec{V}}(\mathscr{S}) = 2\mu \int_{\mathscr{V}} \vec{e}[\vec{u}]:\vec{e}[\vec{v}]\mathrm{d} {\mathscr{V}}.
\end{equation}
We are now ready to differentiate this relation with respect to the variations of the domain $\mathscr{S}$. To this end, we will use the formula for the derivative of integrals on a variable domain, shown in \citet[Theorem 5.2.2]{HENROTPIERRE} in a mathematical setting. Of note, this formula is also a particular application of the so-called Reynolds transport theorem.
\begin{multline}\label{eq:0856}
- \langle dJ_{\vec{V}}(\mathscr{S}),\bm{\theta}\rangle =2\mu \int_{\mathscr{S}} \vec{e}[\vec{u}]:\vec{e}[\vec{v}](\bm{\theta}\vec{\cdot} \vec{n})\mathrm{d} \mathscr{S} \\ +2\mu \int_{\mathscr{V}} \vec{e}[\vec{u'}]:\vec{e}[\vec{v}]\mathrm{d}\mathscr{V}+2\mu \int_{\mathscr{V}} \vec{e}[\vec{u}]:\vec{e}[\vec{v'}]\mathrm{d} {\mathscr{V}},
\end{multline}
where  $(\vec{u'},p')$ and  $(\vec{v'},q')$ may be interpreted as characterising the hypothetical behaviour of the fluid within $\mathscr{B}$ if the surface $\mathscr{S}$ was moving at a speed corresponding to the deformation $\vec{\theta}$. The quantities $(\vec{u'},p')$ and  $(\vec{v'},q')$ are thus solutions of the Stokes-like systems
\begin{equation}
\left \{
\begin{array}{ll}
\mu \Delta \vec{u'} - \nabla p' = \vec{0} & \text{in }\mathscr{V}, \\
\nabla \vec{\cdot} \vec{u'} = 0& \text{in }\mathscr{V}, \\
\vec{u'} = -[\nabla (\vec{u}-\vec{U})]\vec{n} (\vec{\theta}\vec{\cdot}\vec{n}) & \text{on } \mathscr{S}, \\
\vec{u'} = \vec{0} & \text{on } \partial \mathscr{B},
\end{array}
\right .
\label{eq:stokesBisprime}
\end{equation}
and
\begin{equation}
\left \{
\begin{array}{ll}
\mu \Delta \vec{v'} - \nabla q' = \vec{0} & \text{in }\mathscr{V} \\
\nabla \vec{\cdot} \vec{v'} = 0& \text{in }\mathscr{V} \\
\vec{v'} = -[\nabla(\vec{v}-\vec{V})]\vec{n} (\vec{\theta}\vec{\cdot}\vec{n}) & \text{on } \mathscr{S}, \\
\vec{v'} = \vec{0} & \text{on } \partial \mathscr{B}.
\end{array}
\right .
\label{eq:stokesAdjprime}
\end{equation}

Let us rewrite the two last terms of the sum in \eqref{eq:0856} under a convenient form for algorithmic issues. By using an integration by parts, one gets
\begin{equation}
2\mu \int_{\mathscr{V}} \vec{e}[\vec{u'}]:\vec{e}[\vec{v}]\mathrm{d}\mathscr{V} = \mu \int_{\mathscr{V}} (-\Delta \vec{v}+\nabla (\nabla \vec{\cdot} \vec{v}))\vec{\cdot} \vec{u'}\mathrm{d}\mathscr{V} +2\mu \int_{\mathscr{S}} \vec{e}[\vec{v}]\vec{n}\vec{\cdot} \vec{u'}\mathrm{d} \mathscr{S}.
\end{equation}
Using the relations contained in Eqs. \eqref{eq:stokesAdj} for $\vec{v}$ and \eqref{eq:stokesBisprime} for $\vec{u}'$ yields 
\begin{eqnarray*}
2\mu \int_{\mathscr{V}} \vec{e}[\vec{u'}]:\vec{e}[\vec{v}]\mathrm{d}\mathscr{V} &=& - \int_{\mathscr{V}} \nabla q\vec{\cdot} \vec{u'}\mathrm{d}\mathscr{V}-2\mu \int_{\mathscr{S}} (\vec{\theta}\vec{\cdot}\vec{n}) \vec{e}[\vec{v}]\vec{n}\vec{\cdot} \nabla (\vec{u}-\vec{U})\vec{n}  \mathrm{d} \mathscr{S} , \\
&=& - \int_{\mathscr{S}} q \vec{u'}\vec{\cdot} \vec{n}\mathrm{d}\mathscr{V}-2\mu \int_{\mathscr{S}} (\vec{\theta}\vec{\cdot}\vec{n}) \vec{e}[\vec{v}]\vec{n}\vec{\cdot} \nabla (\vec{u}-\vec{U})\vec{n}  \mathrm{d} \mathscr{S} ,
\end{eqnarray*}
which finally leads to
\begin{equation}
2\mu \int_{\mathscr{V}} \vec{e}[\vec{u'}]:\vec{e}[\vec{v}]\mathrm{d}\mathscr{V} =  -\int_{\mathscr{S}} (\vec{\theta}\vec{\cdot}\vec{n}) \vec{\sigma}[\vec{v},q]\vec{n}\vec{\cdot} \nabla (\vec{u}-\vec{U})\vec{n}  \mathrm{d} \mathscr{S} .
\label{eq:prop1-0}
\end{equation}

Now, observe that since $\vec{u}-\vec{U}$ vanishes on $\mathscr{S}$ and is divergence free, and defining the derivative with respect to the normal by $\frac{\partial}{\partial \vec{n}} x_i = \frac{\partial x_i}{\partial x_j} n_j$, one has
\begin{eqnarray*}
\vec{n}\vec{\cdot} \nabla (\vec{u}-\vec{U})\vec{n}&=&\frac{\partial (u_i-U_{i})}{\partial x_j}n_j n_i=\frac{\partial (u_i-U_{i})}{\partial \vec{n}}n_i\\
&=& \frac{\partial (u_i-U_{i})}{\partial x_i}=\nabla\vec{\cdot} (\vec{u}-\vec{U})=0
\end{eqnarray*}
on $\mathscr{S}$, since $\vec{n}\vec{\cdot} \vec{n}=1$ and therefore,
\begin{equation}
\vec{\sigma}[\vec{v},q]\vec{n}\vec{\cdot} \nabla (\vec{u}-\vec{U})\vec{n}=2\mu \vec{e}[\vec{v}]\vec{n}\vec{\cdot} \nabla (\vec{u}-\vec{U})\vec{n}\quad \text{on }\mathscr{S}.
    \label{eq:prop1-1}
\end{equation}
Using straightforward calculations as carried in \citet[Lemma~1]{MR4269970}, we can moreover show that
\begin{equation}
    \vec{e} [\vec{v}] \vec{n} \vec{\cdot} \nabla (\vec{u}-\vec{U}) \vec{n} = \vec{e} [\vec{v}] : \vec{e} [\vec{u}-\vec{U}],
    \label{eq:prop1-2}
\end{equation}
yielding a more symmetrical expression for \eqref{eq:prop1-1}:
$$
\vec{\sigma}[\vec{v},q]\vec{n}\vec{\cdot} \nabla (\vec{u}-\vec{U})\vec{n}=2\mu \vec{e}[\vec{v}]\vec{n}\vec{\cdot} \vec{e}[\vec{u}-\vec{U}]\vec{n}\quad \text{on }\mathscr{S}.
$$
It follows that \eqref{eq:prop1-0} can be rewritten as
\begin{equation}
2\mu \int_{\mathscr{V}} \vec{e}[\vec{u'}]:\vec{e}[\vec{v}]\mathrm{d}\mathscr{V}=-2\mu \int_{\mathscr{S}} (\vec{\theta}\vec{\cdot}\vec{n}) \vec{e}[\vec{v}]: \vec{e} [\vec{u}-\vec{U}] \mathrm{d} \mathscr{S}  .
\label{eq:prop1-3}
\end{equation}
By mimicking the computation above, we obtain similarly
\begin{equation}
2\mu \int_{\mathscr{V}} \vec{e}[\vec{u}]:\vec{e}[\vec{v'}]\mathrm{d}\mathscr{V}=-2\mu  \int_{\mathscr{S}} (\vec{\theta}\vec{\cdot}\vec{n}) \vec{e}[\vec{u}]: \vec{e} [\vec{v}-\vec{V}] \mathrm{d} \mathscr{S}  .
\label{eq:prop1-4}
\end{equation}
Gathering \eqref{eq:0856}, \eqref{eq:prop1-3} and \eqref{eq:prop1-4} yields
$$
-\langle dJ_{\vec{V}}(\mathscr{S}),\bm{\theta}\rangle = 2\mu \int_{\mathscr{S}} (\vec{\theta}\vec{\cdot}\vec{n}) \left(\vec{e}[\vec{u}]:\vec{e}[\vec{v}]-\vec{e}[\vec{v}]: \vec{e} [\vec{u}-\vec{U}]-\vec{e}[\vec{u}]: \vec{e} [\vec{v}-\vec{V}]\right) \mathrm{d} \mathscr{S}  ,
$$ 
and rearranging the terms finally leads to the expected expression of the shape derivative \eqref{eq:shape-diff} and concludes the proof of Proposition \ref{prop:diff}.

As explained above, using the shape gradient formula, one can infer a descent direction that is guaranteed to decrease the objective function, although this is valid only at first order, and therefore as long as the domain variation remains small enough. Hence, to reach an optimal shape in practice, one needs to iterate several times the process of applying small deformations and calculating the new shape gradient on the deformed shape. 

\subsection{Descent direction}\label{sec:descent}

In this section, we focus on how to prescribe the descent direction $\vec{\theta}$ from \eqref{eq:shape_grad}. As described in the previous section, the most natural idea consists in choosing $\vec{\theta} = -G \vec{n}$, ensuring that a small domain variation in this direction decreases the objective function. However, this simple choice can yield vector fields that are not smooth enough, typically leading to numerical instability \citep[see e.g.][]{MR2340012}. To address this issue, a classical method consists in using a variational formulation involving the derivative of $\vec{\theta}$. More precisely, we want to find a field $\vec{\theta}$ that satisfies the following identity \textit{for all} $\vec{\psi} \in \vec{\Theta}$:
\begin{equation}
     \int_{\mathscr{V}}{\nabla \bm{\theta}: \nabla \bm{\psi}\: \mathrm{d} \mathscr{V}} =-\langle dJ(\mathscr{S}),\vec{\psi}\rangle.
     \label{eq:descent1}
\end{equation} 
In particular, evaluating this identity at $\vec{\theta}$ yields
$$
\langle dJ(\mathscr{S}),\bm{\theta}\rangle=- \int_{\mathscr{V}}|{\nabla \bm{\theta}}|^2\mathrm{d}\mathscr{V}\leq 0,
$$
guaranteeing decrease of $J_{\vec{V}}$. Thus, the variational formulation of Equation \eqref{eq:descent1} implicitly defines a good descent direction. To determine $\vec{\theta}$ explicitly, let us now apply Green's formula on Equation \eqref{eq:descent1}:
\begin{equation}
    -\int_{\mathscr{V}} \vec{\psi} \vec{\cdot}  \Delta \vec{\theta} \mathrm{d} \mathscr{V} 
    +\int_{\mathscr{S}} \vec{\psi} \vec{\cdot} ( \nabla \vec{\theta} \vec{n} ) \mathrm{d} \mathscr{S}
    = - \int_{\mathscr{S}} \vec{\psi} \vec{\cdot} G \vec{n} \mathrm{d} \mathscr{S}.
\end{equation}
This identity being valid for all $\vec{\psi}$, we straightforwardly deduce that $\vec{\theta}$ is solution of the Laplace equation
\begin{equation}
  \left\{
  \begin{array}{cl}
    -\Delta \bm{\theta} = \vec{0} & \text{in } \mathscr{V},\\
    \bm{\theta} = \vec{0} & \text{on } \partial\mathscr{B},\\
   {(\nabla \bm{\theta})} \vec{\cdot} \vec{n}= -G \vec{n} & \text{on } \mathscr{S}.
  \end{array}
  \right.
  \label{eq:laplace}
\end{equation}
Note that the dependence of this problem in the criterion $J_{\vec{V}}$ and shape derivative is contained within the boundary condition on $\mathscr{S}$, in which the shape gradient $G$ appears. 

Of note, while this variational method allows to infer a ``good'' descent direction, it is also numerically more costly, since it requires the resolution of the PDE system \eqref{eq:laplace} at every iteration.

Overall, we have shown in this section that the shape derivative for the optimisation of any entry of the grand resistance tensor comes down to a single formula \eqref{eq:shape_grad}, which depends on the solutions to two appropriately chosen resistance problems. In the next section, we see how to numerically apply this theoretical framework to compute various optimised shapes.

\section{Numerical implementation}

In this section, moving further towards a practical use of the shape gradient calculation for the resistance problem, we will introduce a dedicated algorithm of shape optimisation. As described in the previous section, the main idea of this algorithm consists in computing a descent direction $\vec{\theta}$ and applying it to deform ``a little'' the surface of the shape, then iterating these little deformations. The objective function will be monitored to ensure that it gradually decreases in the process and, hopefully, that it eventually converges to a given value, suggesting that the corresponding shape represents a local optimum.



\subsection{Manufacturing constraints}

In addition to this main feature of the algorithm, we typically need to include so-called \textit{manufacturing} constraints on the shape to prevent it from reaching trivial (shrunk to a single point or expanded to fill the entire fluid domain) or unsuitable (e.g. too thin or too irregular) solutions.  As mentioned at the beginning of section \ref{sec:theoFramSO}, in this paper, we chose to focus on the simple and arguably canonical constraint of a constant volume $| \mathscr{V} |$ enclosed by the shape $\mathscr{S}$. Hence, denoting by $V_0$ the volume of the initial solid, we are considering the \textit{constrained} optimisation problem

\begin{equation}\label{SOP}
\max_{| \mathscr{V} | = V_0} J_{\vec{V}} (\mathscr{S}).
\end{equation}

The volume constraint may be enforced with a range of classical optimisation techniques, among which we will use a so-called ``augmented Lagrangian'', adapted from \citet[Section~3.7]{MR3878725} and briefly described in this section. The augmented Lagrangian algorithm converts the constrained optimisation problem \eqref{SOP} into a sequence of unconstrained problems (hereafter indexed by $n$). Hence, we will be led to solve:
\begin{equation}\label{eq.optLn}
 \inf_{\mathscr{S}}{\mathcal{L}(\mathscr{S},\ell^n, b^n)},
\end{equation}
where
\begin{equation}\label{eq.auglag}
  \mathcal{L}(\mathscr{S},\ell,b) = J(\mathscr{S}) - \ell (|\mathscr{V}|-V_0) + \frac{b}{2} (|\mathscr{V}|-V_0)^2.
\end{equation}

In this definition, the parameter $b$ is a (positive) penalisation factor preventing the equality constraint `$|\mathscr{S}|=V_0$' to be violated. The parameter $\ell$ is a Lagrange multiplier associated with this constraint.

The principle of the augmented Lagrangian algorithm rests upon the search for a (local) minimiser $S^n$ of $S \mapsto \mathcal{L}(S, \ell^n,b^n)$
for fixed values of $\ell^n$ and $b^n$. Given $\alpha>1$, these parameters are updated according to the rule:
\begin{equation}\label{eq.uplag}
\ell^{n+1} = \ell^n - b^n (|\mathscr{V}|^n-V_0) , \text{ and } b^{n+1} = \left\{
\begin{array}{cl}
  \alpha b^n & \text{if } b < b_{\text{target}},\\
b^n & \text{otherwise};
\end{array}
\right.
\end{equation}
in other terms, the penalisation parameter $b$ is increased during the first iterations until the value $b_{\text{target}}$ is reached. This regular increase of $b$
ensures that the domain satisfies the constraint more and more precisely during the optimisation process.

\subsection{Numerical resolution of the PDEs}

For the sake of clarity and replicability of the algorithm described below, we provide some additional information about the numerical resolution of the Stokes and Laplace equations (\eqref{eq:stokes}, \eqref{eq:laplace}) required at each step of the deformation. 

The surface $\mathscr{S}$ is first equipped with a triangular surface mesh ${\mathcal T}$ containing the coordinates of the nodes, the middle of the edges, the center of the elements, and connectivity matrices.

The numerical resolution is then carried out by boundary element method \citep{pozrikidis1992boundary} using the BEMLIB Fortran library \citep{pozrikidis2002practical}, which allows to determine the force distribution at each point $\vec{x}$ of the (discretised) surface $\mathscr{S}$ by making use of the integral representation
\begin{equation}
    \vec{u}(\vec{x}) = \int_\mathscr{S}\vec{G}(\vec{x} - \vec{x}_0) \vec{f}(\vec{x}_0) \mathrm{d} \vec{x}_0,
\end{equation}
where $\vec{G}$ is the Oseen tensor given by
\begin{equation}
    G_{ij}(\vec{x}) = \frac{1}{\| \vec{x} \|} \delta_{ij} + \frac{1}{\| \vec{x} \|^3} x_i x_j.
\end{equation}

Once the force distribution $\vec{f}$ is known, the rate-of-strain tensors $\vec{e}$ needed to compute the shape gradient established in formula \eqref{eq:shape-grad} can be conveniently computed through the integral expression

\begin{equation}
    e_{ij} (\vec{x}) = \int_\mathscr{S}\left ( \frac{1}{\| \vec{x} -\vec{x}_0 \|^3} \delta_{ij} (\vec{x} -\vec{x}_0)_k - \frac{3}{\| \vec{x} -\vec{x}_0 \|^5} (\vec{x} -\vec{x}_0)_i (\vec{x} -\vec{x}_0)_j (\vec{x} -\vec{x}_0)_k \right ) f_k (\vec{x}_0) \mathrm{d} \mathscr{S}.
\end{equation}

\subsection{Shape optimisation algorithm}

Let us now describe the main steps of the algorithm.\\

\begin{enumerate}
\item $\;$ \textbf{Initialisation.}
{
\begin{itemize}
\item Equip the initial shape $\mathscr{S}^0$ with a mesh ${\mathcal T}^0$, as described above.
\item Select initial values for the coefficients $\ell^0$, $b^0>0$ of the augmented Lagrangian algorithm.\\
\end{itemize}
}
\item $\;$ \textbf{Main loop: for $n=0, ...$}
{
  \begin{enumerate}
  \item $\;$Compute the solution $(\vec{u}^n,p^n)$ to the Stokes system \eqref{eq:stokes} on the mesh ${\mathcal T}^n$ of $\mathscr{S}^n$;
  \item $\;$Compute the solution $(\vec{v}^n, q^n)$ {to} the adjoint system \eqref{eq:stokesAdj} on the mesh ${\mathcal T}^n$ of $\mathscr{S}^n$.
  \item $\;$Compute the $L^2(\mathscr{S}^n)$ shape gradient $G^n$ of $J$, as well as the shape gradient $\phi^n$ of $\mathscr{S} \mapsto \mathcal{L}(\mathscr{S},\ell^n,b^n)$ given by
  $$
  \phi^n=G^n-\ell^n+b^n(|\mathscr{V}|-V_0).
  $$
  \item $\;$ Infer a descent direction $\bm{\theta}^n$ for $\mathscr{S} \mapsto \mathcal{L}(\mathscr{S},\ell^n,b^n)$ by solving the PDE
\begin{equation}\label{eq:laper}
  \left\{
  \begin{array}{cl}
    -\Delta \bm{\theta} = \vec{0} & \text{in } \mathscr{V}^n,\\
    \bm{\theta} = \vec{0} & \text{on } \partial\mathscr{B},\\
 {(\nabla \bm{\theta})}\vec{n}= -\phi^n \vec{n} & \text{on } \mathscr{S}^n.
  \end{array}
  \right.
\end{equation}
on the mesh ${\mathcal T}^n$.
\item $\;$ \label{step:descent} Find a descent step $\tau^n$ such that
  \begin{equation}\label{eq.declag}
\mathcal{L}((\text{\rm Id}+\tau^n\bm{\theta}^n)(\mathscr{S}^n),\ell^n,b^n) < \mathcal{L}(\mathscr{S}^n,\ell^n,b^n).
  \end{equation}
 \item $\;$ Move the vertices of ${\mathcal T}^n$ according to $\tau^n$ and $\bm{\theta}^n$:
  \begin{equation}\label{eq.xinp1}
    \vec{x}_p^{n+1} = \vec{x}_p^n + \tau^n \bm{\theta}^n(\vec{x}_p^n).
  \end{equation}
 \begin{itemize}
 \item If the resulting mesh is invalid, go back to step \ref{step:descent}, and use a smaller value for $\tau^n$,
 \item Else, the positions (\ref{eq.xinp1}) define the vertices of the new mesh ${\mathcal T}^{n+1}$.
 \end{itemize}
 \item $\;$ If the quality of ${\mathcal T}^{n+1}$ is too low, use a local remeshing.
\item $\;$ Update the augmented Lagrangian parameters according to (\ref{eq.uplag}).\\
  \end{enumerate}
\item $\;$ \textbf{Ending criterion.} Stop if
  \begin{equation}
    \|\bm{\theta}^n\|_{L^2(S^n)} < \varepsilon_{\text{\rm stop}}.
    \label{eq:end}
  \end{equation}
  \textbf{Return} $\mathscr{S}^n$.
  }
  \end{enumerate}

\section{Numerical results}
\label{sec:numerics}

In this section, we present various applications of the algorithm with different entries of the resistance tensor as objective functions. As mentioned above, the initial shape can be chosen freely, but we have made the choice to focus on the symmetric sphere for the initial shape, allowing for easier interpretability. 

An important preliminary remark to these results is that they did not all reach the stopping criterion \eqref{eq:end}. Instead, the algorithm was stopped due to overlapping of the shape or other problem-specific considerations, discussed below. While the displayed shapes are not all strictly local optima for said reasons, the interpretation of the deformation tendencies seem to have great importance from the point of view of fluid mechanics, giving crucial information on the general, ideal characteristics of shapes that offer high or low resistance for a particular entry of the resistance tensor and offering a theoretical backup to previous phenomenological observations.

\begin{figure}
    \centering
    \begin{overpic}[width=\textwidth]{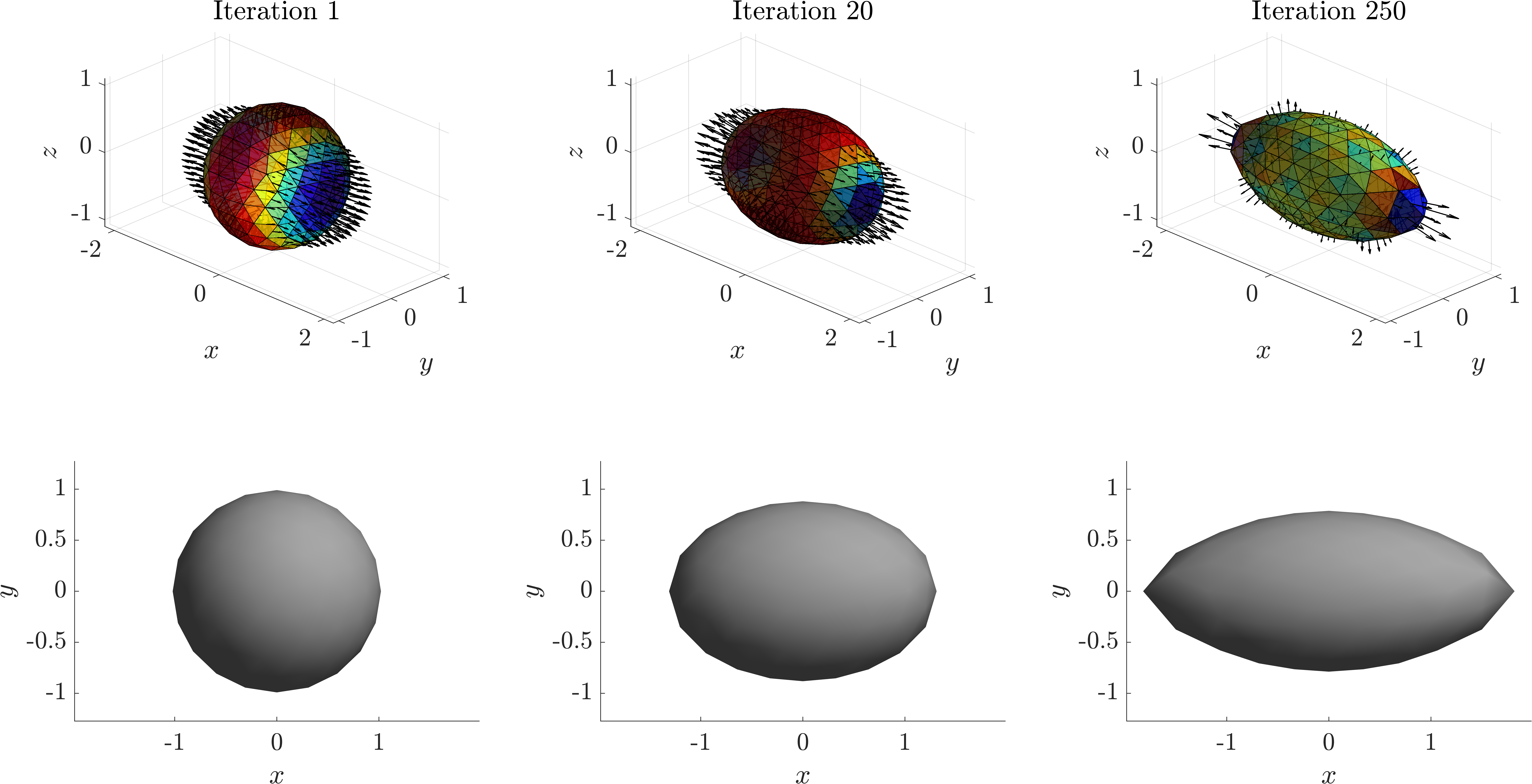}
    \put(4,50){(a)}
    \put(36,50){(b)}
    \put(70,50){(c)}
    \end{overpic}
    \caption{Visualisation of the shape optimisation algorithm running through the minimisation of $K_{11}$. The three columns on the left show the aspect of the shape at three different stages: (a) spherical shape at the first iteration; (b) after 20 iterations; and (c) at the end of the simulation. The surface colours on the top row represent the shape gradient value (from red for a high value for inwards deformation to blue for high outwards deformation), while the arrows show the deformation field $\vec{\theta}$ (normalised for better visualisation). The bottom row allows to observe the evolution of the shape profile, with the final shape closely resembling the well-known optimal drag profile first described by \citet{pironneau1973optimum}.} 
        \label{fig:k11}
\end{figure}

\begin{figure}
        \centering
        \includegraphics[width=0.7\textwidth]{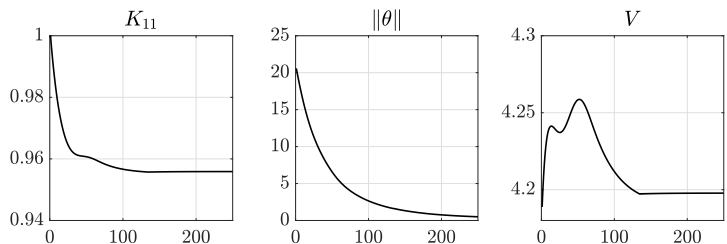}
        \caption{Evolution of $K_{11}$, $\|\vec{\theta}\|$ and the shape volume $V$ along the simulation displayed on \Cref{fig:k11}, strongly suggesting convergence to a minimum of the shape functional.}
        \label{fig:k11_param}
\end{figure}
    
\begin{figure}
    \centering
    \includegraphics[width=\textwidth]{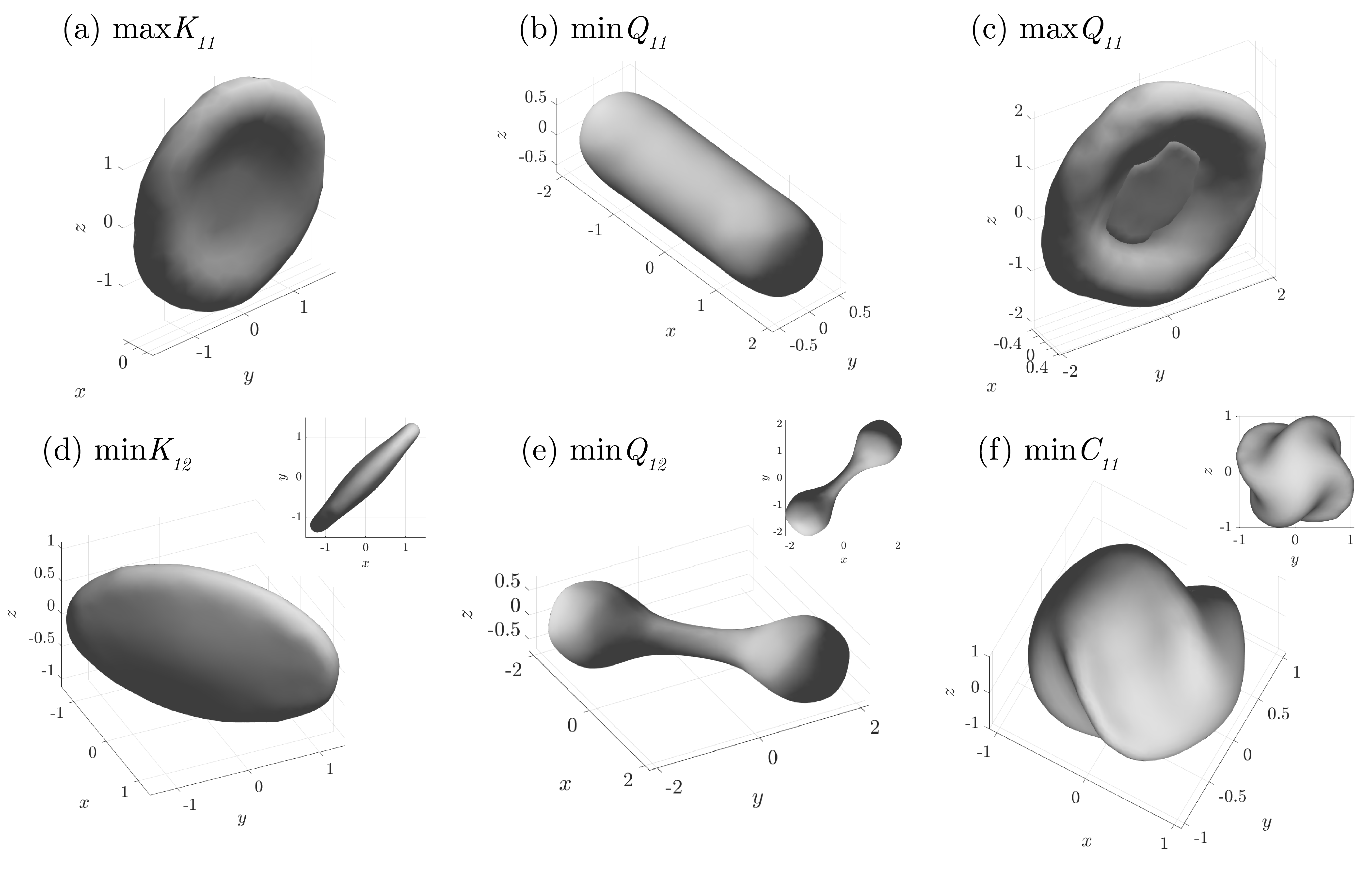}
    \caption{Results obtained from a spherical initial shape and various objective coefficients.}
    \label{fig:shapes}
\end{figure}

\subsection{Diagonal parameters}

The first numerical example will be the resolution of the classical ``minimal drag'' problem at constant volume, equivalent to the minimisation of $K_{11}$. The solution to this problem was determined to be an axisymmetric ``rugby-ball''-like shape by \citet{pironneau1973optimum}, and has been found again with different methods as well as used as an exemplar in an extensive literature later on. We will also use it as a convenient way to check the validity and performance of the algorithm described in the previous section.

The results are shown on figure \ref{fig:k11}. Starting from a sphere, the shape gradient $G$ and deformation field $\vec{\theta}$ are represented on the top left plot (a), with the red and blue colours being respectively associated to positive and negative gradient, meaning inward and outward associated deformation. As expected, the deformation vector field tends to stretch the sphere in the $x$ direction in order to decrease its drag. After 20 iterations (b), the object has taken the shape of an ellipsoid. Of note, axisymmetry, known as a feature of the optimal shape for this problem, is remarkably well preserved along the numerical resolution. At 250 iterations, the ending criterion \eqref{eq:end} is reached and the algorithm stops, with the resulting shape resembling closely the famous ``rugby ball'' of \citet{pironneau1973optimum}. The final drag coefficient is equal to 0.9558, in good agreement with the value known to be the one associated to optimal drag (approx. 0.95425). The small difference between these is attributable to the coarse meshing used for this simulation, which decreases overall precision. Note nonetheless that it is numerically rather remarkable that a mesh of this low quality is able hold the full optimisation problem with good accuracy, suggesting the optimisation framework laid out in this paper enjoys a good level of robustness to coarse discretisation.

The three plots on the Figure \ref{fig:k11_param} show the evolution of the criterion $J_{\vec{V}}(\mathscr{S}) = K_{11}$, the $\mathrm{L}^2$-norm of the deformation vector field $\| \vec{\theta} \|$ and the volume $V$ enclosed by $\mathscr{S}$ along the simulation, with a clear numerical convergence being observed.
Of note, the value of $K_{11}$ is directly correlated to the volume $V$ of the body, making this particular problem extremely sensitive to volume variations. Unlike for the optimisation problems associated to other entries of the resistance tensor, the augmented Lagrangian algorithm with adaptive step described in the previous section was observed to induce instability and amplifying volume oscillations, even with fine tuning of the parameters $\ell$ and $b$. For that reason, the algorithm was adapted for the results displayed on figure \ref{fig:k11}, empirically setting a fixed deformation step $\tau$ and Lagrange multiplier $\ell$ to obtain stability and convergence. The parameter values used in figure \ref{fig:k11} are $\tau = 10^{-3}$, $\ell = 98.8$, $b_0 = 10$, $b_{\mathrm{target}} = 500$ and $\alpha = 1.03$. 

More generally, a good choice of augmented Lagrangian parameters is critical to observe convergence of the algorithm, and is highly dependent on the nature of the problem, therefore requiring \textit{ad hoc} tuning for each different objective function. 

Now, let us turn to the other entries of $\mathsfbi{R}$. Figure \ref{fig:shapes} gathers the results for six different objective functions. These results are, to the best of the authors' knowledge, fully novel and hold several interesting interpretations.

Panels \ref{fig:shapes}(a) and \ref{fig:shapes}(c) display numerical results obtained for maximising the resistance coefficients $K_{11}$ and $Q_{11}$ -- formally, we can indifferently minimise or maximise the criterion $J$ by simply reversing the sign of the shape gradient in Equation \eqref{eq:shape_grad}. As could be expected, maximising the translational drag through $K_{11}$ has the effect of flattening the sphere along the $yz$-plane. Perhaps less intuitive is that the final shape presents a biconcavity evoking those of a red blood cell, and that similar characteristics are observed when maximising the torque-rotation coupling through $Q_{11}$. Of note, for these two situations, the algorithm was stopped due to overlapping of the surface at the center of the biconcavity, visible on panel (c). 

On the other hand, the shape that minimises $Q_{11}$ can be seen on panel \ref{fig:shapes}b. This time, the rotational drag for the sphere appears to be reduced by stretching the shape along the $x$-axis, until reaching a cylinder-like shape with nearly hemispherical extremities. The final shape strikingly evokes the shape of some bacteria species like \textit{Escherichia coli}, with this observation possibly being an argument backing the importance of motility to explain microorganism morphology, among many other factors \citep{yang2016staying,van2017determinants}.

\subsection{Extradiagonal parameters}

Unlike the diagonal entries $K_{ii}$ and $Q_{ii}$ of the resistance tensor, the extradiagonal entries of the grand resistance tensor are not necessarily positive. In fact, a mirror symmetry in along an appropriately chosen plane will reverse the sign of extradiagonal entries. This observation induces that objects possessing certain planar symmetries have null entries in their resistance tensor; in particular, all the extradiagonal entries of a sphere's resistance tensor are equal to zero. These properties importantly imply that the minimisation and maximisation problems are equivalent when choosing an extradiagonal entry as an objective: one can switch between both by means of an appropriate planar symmetry. 

With this being said, the bottom row of figure \ref{fig:shapes} displays results of the minimisation of extradiagonal coefficients of the resistance tensor. The optimal shape for $K_{12}$ can be seen on figure \ref{fig:shapes}d. This may be interpreted as the shape that realises the best transmission from a force applied to a direction (here along the $x$-axis) to a translation towards a perpendicular direction (here, the $y$-axis). The corresponding optimal shape presents a flattened aspect along the diagonal plane $x=y$, and is slightly thicker at the center than at the edges.

In the case of the optimisation of $Q_{12}$ (figure \ref{fig:shapes}e), an interesting ``dumbbell'' shape emerges. This can be understood when considering that maximising $Q_{12}$ accounts for achieving the best transmission of a torque applied in the $x$ direction to a rotation around the $y$ axis. The algorithm finds that the best way to do this is to separate the mass of the sphere in two smaller parts along a $y=x$ line. Of note, convergence of the criterion was not observed when stopping the simulation here; indeed, with suitable remeshing provided, the algorithm would most likely continue indefinitely to spread the two extremities of the dumbbell further aside.  

Finally, one of the most interesting findings lie in the optimisation of $C_{11}$ coefficient, observable on figure \ref{fig:shapes}f. This parameter accounts for the coupling between torque and translation; hence optimising it means that we are looking for the shape that converts best a rotational effect into directional velocity. Helicoidal shapes are well-known to be capable to achieve this conversion. More generally, $C_{11}$ is nonzero only if the shape possesses some level of chirality. Optimisation of $C_{11}$ was tackled for a particular class of shapes in \citet{keaveny2013optimization}, in the context of magnetic helicoidal swimmers. Considering slender shapes parametrised by a one-dimensional curve, they find that optimal shapes are given by regular helicoidal folding, with additional considerations on its pitch and radius depending on parameters and on the presence of a head. The family of optimal shapes however remained rather restrictive compared to our general setting. Starting from a spherical initial shape, which is notably achiral, we can observe on figure \ref{fig:shapes}f the striking emergence of four helicoidal wings, that tend to sharpen along the simulation. Again, the stopping criterion for the shape displayed on figure \ref{fig:shapes}f occurred because of overlapping of the mesh at the edges, and not because the norm of the deformation field converged to zero. While appropriate handling of the narrow parts of the helix wings may allow to carry on the shape optimisation process and observe further folding of the sphere into a long corkscrew-like shape, the observation itself of chirality emerging out of an achiral initial structure throughout an optimisation process is already an arguably captivating result, echoing the widespread existence and importance of chirality among microswimmers, in particular as a possible mean of producing robust directional locomotion within background flows \citep{wheeler2017use}. 

\section{Discussion and perspectives}
\label{sec:discussion}

In this paper, we have addressed the problem of optimal shapes for the resistance problem in a Stokes flow. Considering the entries of the grand resistance tensor as objective shape functionals to optimise, and using the framework of Hadamard boundary variation, we derived a general formula for the shape gradient, allowing to define the best deformation to apply to any given shape. While this shape optimisation framework is mathematically standard, its usage in the context of microhydrodynamics is limited, mostly circumscribed to the work of \citet{keaveny2013optimization}, and the theoretical results and numerical scheme that we presented here provide a much higher level of generality, both concerning the admissible shapes and the range of objective functions.

After validating the numerical capabilities of the shape optimisation algorithm by comparing the optimal shape for $K_{11}$ to the celebrated result of \citet{pironneau1973optimum}, we systematically investigated the shapes minimising and maximising entries of the resistance tensor. The numerical results reveal fascinating new insights on optimal hydrodynamic resistance. In particular, we obtained an optimal profile for the torque drag ($Q_{11}$), observed the emergence of chiral, helicoidal structure maximising the force/rotation coupling ($C_{11}$), and other intriguing shapes generated when minimising extradiagonal entries.

The potentialities of this framework are not limited to the examples displayed in the numerical results section. With most of the optimisation problems considered here being highly unconstrained and nonconvex, we can safely assume that many local extrema exist, and that a range of different results is likely to be observed for different initial shapes. As discussed above, finer handling of the surface mesh to deal with locally high curvature, sharp edges and cusps, and additional manufacturing constraints to prevent self-overlapping and take other criteria into account, are warranted to pursue this broader exploration. Furthermore, seeing as some of the shapes in figure \ref{fig:shapes} appear to take a torus-like profile from an initial spherical shape, it might be interesting to allow topological modifications of the shape along the optimisation process, which requires different approaches such as the level set method \citep{allaire2007conception}.

As mentioned in Section \ref{sec:theoFramSO}, whilst being beyond the fluid mechanics scope of this paper, mathematical questions also arise from this study, such as formal proof of existence and uniqueness of minimisers for the optimisation problem \eqref{SOPgen-min}. 

In the context of low-Reynolds number hydrodynamics, our results provide novel perspectives to the fundamental problem of optimal hydrodynamic resistance for a rigid body, with the optimisation of some entries of the resistance tensor being performed for the first time. Beyond their theoretical interest, these results could help understand and refine some of the the criteria that are believed to govern the morphology of microscopic bodies \citep{yang2016staying,van2017determinants}. 

Furthermore, the computational structure of the optimisation problem is readily adaptable to more complex objective criteria defined as functions of entries of the grand resistance tensor, which allows to tackle relevant quantities for various applications. A prototypical example would be to seek extremal values for the Bretherton constant $B$ \citep{Bretherton1962}, a geometrical parameter for the renowned Jeffery equations \citep{Jeffery1922} which describe the behaviour of an axisymmetric object in a shear flow. As noted by  \citet{ishimoto2020jeffery}, $B$ can be expressed as a rational function of seven distinct entries of the grand resistance tensor. 
For spheroids, $B$ lies between $-1$ and $1$, but nothing theoretically forbids it from being greater than 1 or smaller than $-1$; yet exhibiting realistic shapes achieving it is notoriously difficult \citep{Bretherton1962,Singh2013}. 
Further, another geometrical parameter $C$ is introduced for chiral helicoidal particles in \citet{ishimoto2020helicoidal} This shape constant, now termed as the Ishimoto constant \citep{ohmura2021near}, characterises the level of chirality and is useful to study bacterial motility in flow \cite{jing2020chirality}
Whilst this parameter can be expressed with respect to entries of the resistance tensor in a similar manner as $B$, very little is known about typical shapes for a given value of $C$, not to mention shapes optimising $C$. The framework developed in this paper provides a promising way of investigating these questions. 

Finally, various refinements of the Stokes problem \ref{eq:stokes} can be fathomed to address other open problems in microhydrodynamics and microswimming. Dirichlet boundary conditions on the object surface, considered in this paper as well as in a vast part of the literature, may fail to properly describe the fluid friction arising at small scale, notably when dealing with complex biological surfaces. Nonstandard boundary conditions such as the Navier conditions \citep{B616490K} are then warranted. Interestingly, the optimal drag problem for a rigid body, although well resolved since long for Dirichlet conditions \citep{pironneau1973optimum}, is still open for Navier conditions. 

Seeking to further connect shape optimisation to efficient swimming at microscale, one could also include some level of deformability of the object, which requires to couple the Stokes equation with an elasticity problem. A simple model in this spirit was recently introduced in the context of shape optimisation in \citet{calisti2021synthesis}. Another problem with biological relevance it the optimisation of hydrodynamic resistance when interacting with a more or less complex environment, such as a neigbouring wall or a channel, which is known to change locomotion strategies for microorganisms \citep{elgeti2016microswimmers}; overall, a dynamical, environment-sensitive shape optimisation study stemming from this paper's framework could provide key insights on microswimming and microrobot design. 


\backsection[Funding]{C.M. is a JSPS Postdoctoral Fellow (P22023), and acknowledges partial support by the Research Institute for Mathematical Sciences, an International Joint Usage/Research Center located at Kyoto University and JSPS-KAKENHI Grant-in Aid for JSPS Fellows (Grant No. 22F22023). K.I. acknowledges JSPS-KAKENHI for Young Researchers (Grant No. 18K13456), JSPS-KAKENHI for Transformative Research Areas (Grant No. 21H05309) and JST, PRESTO, Japan (Grant No. JPMJPR1921). Y.P. was partially supported by the ANR Project VirtualChest ANR-16-CE19-0014.}

\backsection[Declaration of interests]{The authors report no conflict of interest.}


\backsection[Author ORCID]{C. Moreau, https://orcid.org/0000-0002-8557-1149; K. Ishimoto, https://orcid.org/0000-0003-1900-7643; Y. Privat, https://orcid.org/0000-0002-2039-7223.}


\appendix

\section{Complements on shape optimisation theory}\label{append:diff}

We need to mind the regularity at three levels of the shape optimisation problem: the shape itself, the deformation vector field, and the solution of the Stokes equations whose the criterion $J_{\vec{V}}$ depends on.

\subsection{Well-posedness of the Stokes equation} 

For an open set $\mathscr{V}$ of $\mathbb{R}^3$, we let $H^k(\mathscr{V})$ denote the Sobolev space of functions $v\in L^2(\mathscr{V})$ such that for every multi-index $\alpha$ with $|\alpha|\leq k$, its $\alpha$-th derivative of the sense of distributions belongs to $L^2(\mathscr{V})$. We refer to \citet[Chapter~4]{GALDI}, and in particular the theorems IV.1.1 and IV.5.1.

We comment here on the existence, uniqueness and regularity of solutions for System~\eqref{eq:stokes}. Let us assume that $\partial\Omega$ is of class $\mathscr{C}^2$. Under the compatibility conditions
\begin{equation}\label{eq:comp}
    \int_{\mathscr{S}}\vec{U}\cdot \vec{n} \mathrm{d} \mathscr{S}+\int_{\Gamma}\vec{U^\infty}\cdot \vec{n} \mathrm{d} \Gamma
\end{equation}
where $\Gamma=\partial\mathscr{B}$, System~\eqref{eq:stokes} has a unique solution $(\vec{u},p)$ belonging moreover to  $[H^2(\mathscr{V})]^3\times H^1(\mathscr{V})$.

Finally, it can be observed that \eqref{eq:comp} is automatically satisfied for the boundary data $(\vec{U},\vec{U^\infty}$ defined by \eqref{def:U} and \eqref{def:Uinfty}. Indeed, by using the divergence theorem, one has
$$
\int_{\Gamma}\vec{U^\infty}\cdot \vec{n} \mathrm{d} \Gamma=\int_{\mathscr{B}}\nabla \cdot \vec{U^\infty}\mathrm{d}\mathscr{V}=0.
$$
Indeed, the divergence of the cross product vanishes obviously. Regarding the term $\nabla\cdot (\vec{E^\infty x})$, we conclude by using that the trace of $\vec{E^\infty}$ is equal to zero, since $\vec{E^\infty}$ is the shear flow component of $\vec{U^\infty}$.

The term $\int_{\mathscr{S}}\vec{U}\cdot \vec{n} \mathrm{d} \mathscr{S}$ can be handled similarly.

\subsection{Set of admissible shapes}\label{append:funSpace}

We recall that the Sobolev space $W^{k,\infty}(\mathbb{R}^3,\mathbb{R}^3)$ is defined as the set of all vector fields $f :\mathbb{R}^3\to \mathbb{R}^3$ such that for every multi-index $\alpha$ with $|\alpha| \leq k$, the mixed partial derivative $D^\alpha f$ exists in a distributional sense and belongs to $L^\infty(\mathbb{R}^3,\mathbb{R}^3))$. Equipped with the norm
$$
\Vert f \Vert_{W^{k,\infty}(\mathbb{R}^3,\mathbb{R}^3)}:=\max_{|\alpha|\leq k}\Vert D^\alpha f\Vert_{L^\infty(\mathbb{R}^3,\mathbb{R}^3)},
$$
it defines a Banach space.

In the Hadamard method, the sensitivity of a shape functional is evaluated with respect to small perturbations of its boundary: more precisely, we consider variations of a given domain $\mathscr{S}$ of the form
\begin{equation}\label{eq.varOm}
 \mathscr{S}_{\bm{\theta}} = (\text{\rm Id} + {\bm{\theta}})(\mathscr{S}),
\end{equation}
where $\bm{\theta}:\mathbb{R}^d \to \mathbb{R}^d$ is a `small' vector field, and $\text{\rm Id}$ is the identity mapping from $\mathbb{R}^d$ into itself.

The admissible shapes  $\Omega$ we will consider will be assumed to be smooth, it is natural that the perturbations vector field $\bm{\theta}$ belong to the set $ \Theta_{ad}$  defined by:
$$ \Theta_{ad} = \left\{ \bm{\theta} : \mathbb{R}^d \to \mathbb{R}^d \text{ smooth}, \: \bm{\theta} = 0 \text{ in } \mathbb{R}^d\backslash \overline{\mathscr{B}} \right\};$$
so that variations (\ref{eq.varOm}) of admissible shapes stay admissible.

\subsection{Deformation vector field}

Let us add some technical details about the regularity of the vector field $\vec{\theta}$ to consider. 
Let $\Omega_0$ denote an open bounded subset of $\mathbb{R}^3$ with a $\mathcal{C}^2$ boundary and let $\bm{\theta}$ belong to $W^{3,\infty}(\mathbb{R}^3,\mathbb{R}^3)$ 
and such that
$$
\Vert \bm{\theta}\Vert_{W^{3,\infty}(\mathbb{R}^3,\mathbb{R}^3)} <1.
$$
Then  $(\text{\rm Id} + {\bm{\theta}})(\Omega_0)$ is an open bounded domain whose boundary is of class $\mathcal{C}^{2}$. Furthermore, $\text{\rm Id} + {\bm{\theta}}$ is a diffeomorphism and one has $(\text{\rm Id} + {\bm{\theta}})(\partial\Omega_0)=\partial((\text{\rm Id} + {\bm{\theta}})(\Omega_0))$.

As a consequence, since one aims at dealing with domains having a $\mathcal{C}^2$ boundary, so that solutions of the involved PDEs will be understood in a strong sense, we will deal with vector fields ${\bm{\theta}}$ in
$$ 
\Theta_{ad} = \left\{ \bm{\theta} \in W^{3,\infty}( \mathbb{R}^d, \mathbb{R}^d), \: \bm{\theta} = 0 \text{ in } \mathbb{R}^d\backslash \overline{\mathscr{B}} \right\}.
$$

The remainder term in \eqref{eq:shape-diff} is therefore such that
$$
\frac{o(\bm{\theta})}{\Vert \bm{\theta}\Vert_{W^{3,\infty}(\mathbb{R}^3,\mathbb{R}^3)}} \to 0\quad \text{as } \Vert \bm{\theta}\Vert_{W^{3,\infty}(\mathbb{R}^3,\mathbb{R}^3)}\to 0.
$$






\bibliographystyle{jfm.bst}
\bibliography{biblioshape.bib}

\begin{thebibliography}{48}
\expandafter\ifx\csname natexlab\endcsname\relax\def\natexlab#1{#1}\fi
\def\au#1{#1} \def\ed#1{#1} \def\yr#1{#1}\def\at#1{#1}\def\jt#1{\textit{#1}}
  \def\bt#1{#1}\def\bvol#1{\textbf{#1}} \def\vol#1{#1} \def\pg#1{#1}
  \def\publ#1{#1}\def\arxiv#1{#1}\def\org#1{#1}\def\st#1{\textit{#1}}

\bibitem[Allaire(2007)]{allaire2007conception}
{\sc \au{Allaire, Gr\'{e}goire}} \yr{2007} {\em Conception optimale de
  structures\/},  \st{Math\'{e}matiques \& Applications (Berlin) [Mathematics
  \& Applications]},  \vol{vol.~58}.  \publ{Springer-Verlag, Berlin}, with the
  collaboration of Marc Schoenauer (INRIA) in the writing of Chapter 8.

\bibitem[Allaire {\em et~al.\/}(2004)Allaire, Jouve \& Toader]{ajt}
{\sc \au{Allaire, Gr{\'e}goire}, \au{Jouve, Fran{\c{c}}ois} \& \au{Toader,
  Anca-Maria}} \yr{2004}  \at{Structural optimization using sensitivity
  analysis and a level-set method}.  \jt{Journal of computational physics}
  \bvol{194}~(1),  \pg{363--393}.

\bibitem[Bendsoe \& Sigmund(2013)]{bendsig}
{\sc \au{Bendsoe, Martin~Philip} \& \au{Sigmund, Ole}} \yr{2013} {\em Topology
  optimization: theory, methods, and applications\/}.  \publ{Springer Science
  \& Business Media}.

\bibitem[Berti {\em et~al.\/}(2021)Berti, Binois, Alouges, Aussal, Prud'Homme
  \& Giraldi]{berti2021shapes}
{\sc \au{Berti, Luca}, \au{Binois, Micka{\"e}l}, \au{Alouges, Fran{\c{c}}ois},
  \au{Aussal, Matthieu}, \au{Prud'Homme, Christophe} \& \au{Giraldi, Laetitia}}
  \yr{2021}  \at{Shapes enhancing the propulsion of multiflagellated helical
  microswimmers}.  \jt{arXiv preprint arXiv:2103.05637} .

\bibitem[Bocquet \& Barrat(2007)]{B616490K}
{\sc \au{Bocquet, Lydéric} \& \au{Barrat, Jean-Louis}} \yr{2007}  \at{Flow
  boundary conditions from nano- to micro-scales}.  \jt{Soft Matter}  \bvol{3},
   \pg{685--693}.

\bibitem[Borrvall \& Petersson(2003)]{borvall}
{\sc \au{Borrvall, T.} \& \au{Petersson, J.}} \yr{2003}  \at{Topology
  optimization of fluids in stokes flow}.  \jt{Int. J. Numer. Meth. Fluids}
  \bvol{41},  \pg{77--107}.

\bibitem[Bourot(1974)]{bourot1974numerical}
{\sc \au{Bourot, J-M}} \yr{1974}  \at{On the numerical computation of the
  optimum profile in stokes flow}.  \jt{Journal of Fluid Mechanics}
  \bvol{65}~(3),  \pg{513--515}.

\bibitem[Bretherton(1962)]{Bretherton1962}
{\sc \au{Bretherton, Francis~P}} \yr{1962}  \at{The motion of rigid particles
  in a shear flow at low reynolds number}.  \jt{Journal of Fluid Mechanics}
  \bvol{14}~(2),  \pg{284--304}.

\bibitem[Calisti(2021)]{calisti2021synthesis}
{\sc \au{Calisti, Valentin}} \yr{2021}  \at{Synthesis of microstructures by
  topological optimization, and shape optimization of a fluid structure
  interaction problem}. PhD thesis, Universit{\'e} de Lorraine.

\bibitem[Courtais {\em et~al.\/}(2021)Courtais, Latifi, Lesage \&
  Privat]{MR4269970}
{\sc \au{Courtais, Alexis}, \au{Latifi, Abderrazak~M.}, \au{Lesage,
  Fran\c{c}ois} \& \au{Privat, Yannick}} \yr{2021}  \at{Shape optimization of
  fixed-bed reactors in process engineering}.  \jt{SIAM J. Appl. Math.}
  \bvol{81}~(3),  \pg{1141--1165}.

\bibitem[Daddi-Moussa-Ider {\em et~al.\/}(2021)Daddi-Moussa-Ider, Nasouri,
  Vilfan \& Golestanian]{daddi2021optimal}
{\sc \au{Daddi-Moussa-Ider, Abdallah}, \au{Nasouri, Babak}, \au{Vilfan, Andrej}
  \& \au{Golestanian, Ramin}} \yr{2021}  \at{Optimal swimmers can be pullers,
  pushers or neutral depending on the shape}.  \jt{Journal of Fluid Mechanics}
  \bvol{922}.

\bibitem[Dapogny {\em et~al.\/}(2018)Dapogny, Frey, Omn\`es \&
  Privat]{MR3878725}
{\sc \au{Dapogny, Charles}, \au{Frey, Pascal}, \au{Omn\`es, Florian} \&
  \au{Privat, Yannick}} \yr{2018}  \at{Geometrical shape optimization in fluid
  mechanics using {F}ree{F}em++}.  \jt{Struct. Multidiscip. Optim.}
  \bvol{58}~(6),  \pg{2761--2788}.

\bibitem[Do\v{g}an {\em et~al.\/}(2007)Do\v{g}an, Morin, Nochetto \&
  Verani]{MR2340012}
{\sc \au{Do\v{g}an, G.}, \au{Morin, P.}, \au{Nochetto, R.~H.} \& \au{Verani,
  M.}} \yr{2007}  \at{Discrete gradient flows for shape optimization and
  applications}.  \jt{Comput. Methods Appl. Mech. Engrg.}  \bvol{196}~(37-40),
  \pg{3898--3914}.

\bibitem[Elgeti \& Gompper(2016)]{elgeti2016microswimmers}
{\sc \au{Elgeti, Jens} \& \au{Gompper, Gerhard}} \yr{2016}  \at{Microswimmers
  near surfaces}.  \jt{The European Physical Journal Special Topics}
  \bvol{225}~(11),  \pg{2333--2352}.

\bibitem[Evgrafov(2006)]{evgrafov}
{\sc \au{Evgrafov, Anton}} \yr{2006}  \at{Topology optimization of slightly
  compressible fluids}.  \jt{ZAMM-Journal of Applied Mathematics and
  Mechanics/Zeitschrift f{\"u}r Angewandte Mathematik und Mechanik}
  \bvol{86}~(1),  \pg{46--62}.

\bibitem[Fujita \& Kawai(2001)]{fujita2001optimum}
{\sc \au{Fujita, Tatsuya} \& \au{Kawai, Tatsuo}} \yr{2001}  \at{Optimum shape
  of a flagellated microorganism}.  \jt{JSME International Journal Series C
  Mechanical Systems, Machine Elements and Manufacturing}  \bvol{44}~(4),
  \pg{952--957}.

\bibitem[Galdi(2011)]{GALDI}
{\sc \au{Galdi, G.~P.}} \yr{2011} {\em An introduction to the mathematical
  theory of the Navier-Stokes equations\/}, 2nd edn. {\em Springer Monographs
  in Mathematics\/} .  \publ{Springer, New York}, steady-state problems.

\bibitem[Henrot \& Pierre(2018)]{HENROTPIERRE}
{\sc \au{Henrot, Antoine} \& \au{Pierre, Michel}} \yr{2018} {\em Shape
  variation and optimization\/},  \st{EMS Tracts in Mathematics},
  \vol{vol.~28}.  \publ{European Mathematical Society (EMS), Z\"{u}rich}, a
  geometrical analysis, English version of the French publication [ MR2512810]
  with additions and updates.

\bibitem[Henrot \& Privat(2010)]{MR2601075}
{\sc \au{Henrot, Antoine} \& \au{Privat, Yannick}} \yr{2010}  \at{What is the
  optimal shape of a pipe?}  \jt{Arch. Ration. Mech. Anal.}  \bvol{196}~(1),
  \pg{281--302}.

\bibitem[Ishimoto(2016)]{ishimoto2016hydrodynamic}
{\sc \au{Ishimoto, Kenta}} \yr{2016}  \at{Hydrodynamic evolution of sperm
  swimming: Optimal flagella by a genetic algorithm}.  \jt{Journal of
  Theoretical Biology}  \bvol{399},  \pg{166--174}.

\bibitem[Ishimoto(2020{\natexlab{{\em a\/}}})]{ishimoto2020helicoidal}
{\sc \au{Ishimoto, Kenta}} \yr{2020{\natexlab{{\em a\/}}}}  \at{Helicoidal
  particles and swimmers in a flow at low reynolds number}.  \jt{Journal of
  Fluid Mechanics}  \bvol{892}.

\bibitem[Ishimoto(2020{\natexlab{{\em b\/}}})]{ishimoto2020jeffery}
{\sc \au{Ishimoto, Kenta}} \yr{2020{\natexlab{{\em b\/}}}}  \at{Jeffery orbits
  for an object with discrete rotational symmetry}.  \jt{Physics of Fluids}
  \bvol{32}~(8),  \pg{081904}.

\bibitem[Jeffery(1922)]{Jeffery1922}
{\sc \au{Jeffery, G.~B.}} \yr{1922}  \at{{The motion of ellipsoidal particles
  immersed in a viscous fluid}}.  \jt{Proceedings of the Royal Society A:
  Mathematical, Physical and Engineering Sciences}  \bvol{102}~(715),
  \pg{161--179}.

\bibitem[Jing {\em et~al.\/}(2020)Jing, Z{\"o}ttl, Cl{\'e}ment \&
  Lindner]{jing2020chirality}
{\sc \au{Jing, Guangyin}, \au{Z{\"o}ttl, Andreas}, \au{Cl{\'e}ment, {\'E}ric}
  \& \au{Lindner, Anke}} \yr{2020}  \at{Chirality-induced bacterial rheotaxis
  in bulk shear flows}.  \jt{Science advances}  \bvol{6}~(28),  \pg{eabb2012}.

\bibitem[Keaveny {\em et~al.\/}(2013)Keaveny, Walker \&
  Shelley]{keaveny2013optimization}
{\sc \au{Keaveny, Eric~E}, \au{Walker, Shawn~W} \& \au{Shelley, Michael~J}}
  \yr{2013}  \at{Optimization of chiral structures for microscale propulsion}.
  \jt{Nano letters}  \bvol{13}~(2),  \pg{531--537}.

\bibitem[Kim \& Karrila(2005)]{Kim2005}
{\sc \au{Kim, S} \& \au{Karrila, S~J}} \yr{2005} {\em {Microhydrodynamics:
  Principles and Selected Applications}\/}. {\em Butterworth - Heinemann series
  in chemical engineering\/} .  \publ{Dover Publications}.

\bibitem[Lauga(2020)]{lauga2020fluid}
{\sc \au{Lauga, Eric}} \yr{2020} {\em The fluid dynamics of cell motility\/}, ,
   \vol{vol.~62}.  \publ{Cambridge University Press}.

\bibitem[Mohammadi \& Pironneau(2010)]{MR2567067}
{\sc \au{Mohammadi, Bijan} \& \au{Pironneau, Olivier}} \yr{2010} {\em Applied
  shape optimization for fluids\/}, 2nd edn. {\em Numerical Mathematics and
  Scientific Computation\/} 2567067.  \publ{Oxford University Press, Oxford}.

\bibitem[Montenegro-Johnson \& Lauga(2015)]{montenegro2015other}
{\sc \au{Montenegro-Johnson, Thomas~D} \& \au{Lauga, Eric}} \yr{2015}  \at{The
  other optimal stokes drag profile}.  \jt{Journal of Fluid Mechanics}
  \bvol{762}.

\bibitem[Ohmura {\em et~al.\/}(2021)Ohmura, Nishigami, Taniguchi, Nonaka,
  Ishikawa \& Ichikawa]{ohmura2021near}
{\sc \au{Ohmura, Takuya}, \au{Nishigami, Yukinori}, \au{Taniguchi, Atsushi},
  \au{Nonaka, Shigenori}, \au{Ishikawa, Takuji} \& \au{Ichikawa, Masatoshi}}
  \yr{2021}  \at{Near-wall rheotaxis of the ciliate tetrahymena induced by the
  kinesthetic sensing of cilia}.  \jt{Science Advances}  \bvol{7}~(43),
  \pg{eabi5878}.

\bibitem[Osher \& Sethian(1988)]{osher1988fronts}
{\sc \au{Osher, Stanley} \& \au{Sethian, James~A}} \yr{1988}  \at{Fronts
  propagating with curvature-dependent speed: algorithms based on
  hamilton-jacobi formulations}.  \jt{Journal of computational physics}
  \bvol{79}~(1),  \pg{12--49}.

\bibitem[Pironneau(1973)]{pironneau1973optimum}
{\sc \au{Pironneau, Olivier}} \yr{1973}  \at{On optimum profiles in stokes
  flow}.  \jt{Journal of Fluid Mechanics}  \bvol{59}~(1),  \pg{117--128}.

\bibitem[Pozrikidis(2002)]{pozrikidis2002practical}
{\sc \au{Pozrikidis, Constantine}} \yr{2002} {\em A practical guide to boundary
  element methods with the software library BEMLIB\/}.  \publ{CRC Press}.

\bibitem[Pozrikidis {\em et~al.\/}(1992)]{pozrikidis1992boundary}
{\sc \au{Pozrikidis, Constantine} \& \au{others}} \yr{1992} {\em Boundary
  integral and singularity methods for linearized viscous flow\/}.
  \publ{Cambridge university press}.

\bibitem[Purcell(1977)]{purcell1977life}
{\sc \au{Purcell, Edward~M}} \yr{1977}  \at{Life at low reynolds number}.
  \jt{American journal of physics}  \bvol{45}~(1),  \pg{3--11}.

\bibitem[Quispe {\em et~al.\/}(2019)Quispe, Oulmas \&
  R{\'e}gnier]{quispe2019geometry}
{\sc \au{Quispe, Johan~E}, \au{Oulmas, Ali} \& \au{R{\'e}gnier, St{\'e}phane}}
  \yr{2019} Geometry optimization of helical swimming at low reynolds number.
  \bt{In {\em 2019 International Conference on Manipulation, Automation and
  Robotics at Small Scales (MARSS)\/}},  \pg{pp. 1--6}. IEEE.

\bibitem[Richardson(1995)]{richardson1995optimum}
{\sc \au{Richardson, S}} \yr{1995}  \at{Optimum profiles in two-dimensional
  stokes flow}.  \jt{Proceedings of the Royal Society of London. Series A:
  Mathematical and Physical Sciences}  \bvol{450}~(1940),  \pg{603--622}.

\bibitem[Ryabov {\em et~al.\/}(2021)Ryabov, Kerimoglu, Litchman, Olenina,
  Roselli, Basset, Stanca \& Blasius]{ryabov2021shape}
{\sc \au{Ryabov, Alexey}, \au{Kerimoglu, Onur}, \au{Litchman, Elena},
  \au{Olenina, Irina}, \au{Roselli, Leonilde}, \au{Basset, Alberto},
  \au{Stanca, Elena} \& \au{Blasius, Bernd}} \yr{2021}  \at{Shape matters: the
  relationship between cell geometry and diversity in phytoplankton}.
  \jt{Ecology letters}  \bvol{24}~(4),  \pg{847--861}.

\bibitem[Sethian \& Wiegmann(2000)]{sethian2000structural}
{\sc \au{Sethian, James~A} \& \au{Wiegmann, Andreas}} \yr{2000}  \at{Structural
  boundary design via level set and immersed interface methods}.  \jt{Journal
  of computational physics}  \bvol{163}~(2),  \pg{489--528}.

\bibitem[Singh {\em et~al.\/}(2013)Singh, Koch \& Stroock]{Singh2013}
{\sc \au{Singh, Vikram}, \au{Koch, Donald~L} \& \au{Stroock, Abraham~D}}
  \yr{2013}  \at{Rigid ring-shaped particles that align in simple shear flow}.
  \jt{Journal of Fluid Mechanics}  \bvol{722},  \pg{121--158}.

\bibitem[Srivastava(2011)]{srivastava2011optimum}
{\sc \au{Srivastava, Deepak~Kumar}} \yr{2011}  \at{Optimum cross section
  profile in axisymmetric stokes flow}.  \jt{Journal of fluids engineering}
  \bvol{133}~(10).

\bibitem[van Teeseling {\em et~al.\/}(2017)van Teeseling, de~Pedro \&
  Cava]{van2017determinants}
{\sc \au{van Teeseling, Muriel~CF}, \au{de~Pedro, Miguel~A} \& \au{Cava,
  Felipe}} \yr{2017}  \at{Determinants of bacterial morphology: from
  fundamentals to possibilities for antimicrobial targeting}.  \jt{Frontiers in
  microbiology}  \bvol{8},  \pg{1264}.

\bibitem[Vilfan(2012)]{vilfan2012optimal}
{\sc \au{Vilfan, Andrej}} \yr{2012}  \at{Optimal shapes of surface slip driven
  self-propelled microswimmers}.  \jt{Physical review letters}
  \bvol{109}~(12),  \pg{128105}.

\bibitem[{Walker} \& {Keaveny}(2013)]{Walker2013}
{\sc \au{{Walker}, Shawn~W.} \& \au{{Keaveny}, Eric~E.}} \yr{2013}
  \at{{Analysis of shape optimization for magnetic microswimmers}}.  \jt{{SIAM
  J. Control Optim.}}  \bvol{51}~(4),  \pg{3093--3126}.

\bibitem[Wang {\em et~al.\/}(2003)Wang, Wang \& Guo]{wang}
{\sc \au{Wang, Michael~Yu}, \au{Wang, Xiaoming} \& \au{Guo, Dongming}}
  \yr{2003}  \at{A level set method for structural topology optimization}.
  \jt{Computer methods in applied mechanics and engineering}  \bvol{192}~(1),
  \pg{227--246}.

\bibitem[Wheeler(2017)]{wheeler2017use}
{\sc \au{Wheeler, Richard~John}} \yr{2017}  \at{Use of chiral cell shape to
  ensure highly directional swimming in trypanosomes}.  \jt{PLoS computational
  biology}  \bvol{13}~(1),  \pg{e1005353}.

\bibitem[Yang {\em et~al.\/}(2016)Yang, Blair \& Salama]{yang2016staying}
{\sc \au{Yang, Desir{\'e}e~C}, \au{Blair, Kris~M} \& \au{Salama, Nina~R}}
  \yr{2016}  \at{Staying in shape: the impact of cell shape on bacterial
  survival in diverse environments}.  \jt{Microbiology and Molecular Biology
  Reviews}  \bvol{80}~(1),  \pg{187--203}.

\bibitem[Zabarankin(2013)]{zabarankin2013minimum}
{\sc \au{Zabarankin, Michael}} \yr{2013}  \at{Minimum-resistance shapes in
  linear continuum mechanics}.  \jt{Proceedings of the Royal Society A:
  Mathematical, Physical and Engineering Sciences}  \bvol{469}~(2160),
  \pg{20130206}.

\end{thebibliography}

\end{document}